\documentclass[twocolumn,amsmath,amssymb,superscriptaddress,aps,prl,floatfix,footinbib,longbibliography]{revtex4-2}
\usepackage[]{graphicx}
\usepackage{tabularx}
\usepackage[usenames,dvipsnames]{color}
\usepackage{bm}
\usepackage{booktabs}
\usepackage{times}
\usepackage{amsfonts}
\usepackage{mathrsfs}
\usepackage{graphicx}%
\usepackage{dcolumn}%
\usepackage{bm}%
\usepackage[dvipsnames]{xcolor}

\usepackage[colorlinks,bookmarks=false,citecolor=blue,linkcolor=red,urlcolor=blue]{hyperref}
\usepackage{multirow}
\usepackage{physics}

\usepackage{mathtools}
\usepackage{makecell}
\usepackage{subcaption}
\usepackage{placeins}
\begin{document}

\title{Orbital Hall Effect in Weyl Semimetals from quantum geometric band interference}

\author{Chiara Pacella}
\affiliation{Department of Physics and Astronomy, University of Bologna, 40127 Bologna, Italy}\email{chiara.pacella2@unibo.it}
\affiliation{Max Planck Institute for the Structure and Dynamics of Matter, Hamburg, Germany}

\author{Maximilian \"Unzelmann}\email{maximilian.uenzelmann@uni-wuerzburg.de}\affiliation{Experimentelle Physik VII and W\"urzburg-Dresden Cluster of Excellence ctd.qmat, Universit\"at W\"urzburg, Am Hubland, D-97074 W\"urzburg, Germany}

\author{Ahmed Osman}\affiliation{Department of Physics and Astronomy, University of Bologna, 40127 Bologna, Italy}\email{ahmed.osman2@studio.unibo.it}

\author{Tim Figgemeier}\affiliation{Experimentelle Physik VII and W\"urzburg-Dresden Cluster of Excellence ctd.qmat, Universit\"at W\"urzburg, Am Hubland, D-97074 W\"urzburg, Germany}

\author{Friedrich Reinert}\affiliation{Experimentelle Physik VII and W\"urzburg-Dresden Cluster of Excellence ctd.qmat, Universit\"at W\"urzburg, Am Hubland, D-97074 W\"urzburg, Germany}

\author{Angel Rubio}
\affiliation{Max Planck Institute for the Structure and Dynamics of Matter, Hamburg, Germany}
\affiliation{Nano-Bio Spectroscopy Group, Departmento de Física de Materiales, Universidad del País Vasco, San Sebastián, Spain}
\affiliation{Initiative for Computational Catalysis (ICC) and Center for Computational Quantum Physics (CCQ), The Flatiron Institute, New York, NY, USA}

\author{Domenico Di Sante}
\affiliation{Department of Physics and Astronomy, University of Bologna, 40127 Bologna, Italy}\email{domenico.disante@unibo.it}

\date{\today}

\begin{abstract}
Orbital angular momentum (OAM) transport in solids, prominently manifested in the orbital Hall effect, has emerged as a fundamental phenomenon that can decisively exceed its spin-based counterparts. However, the microscopic mechanisms governing OAM dynamics remain only partially understood. In particular, the role of band geometry in orbital transport is still largely unresolved. Here we address this question in the TaAs family of Weyl semimetals, TaAs, TaP, NbAs, and NbP, whose well-established topology and associated OAM textures make them an ideal platform in this context. Using ab initio density functional theory, complemented by a minimal Weyl model based on adiabatic perturbation theory, we establish --- both numerically and analytically --- a direct link between OAM transport, band geometry, and topological electronic structure.
\end{abstract}

\maketitle

\begin{figure*}[t!h!]
\centering
\includegraphics[width=0.8\textwidth,angle=0,clip=true]{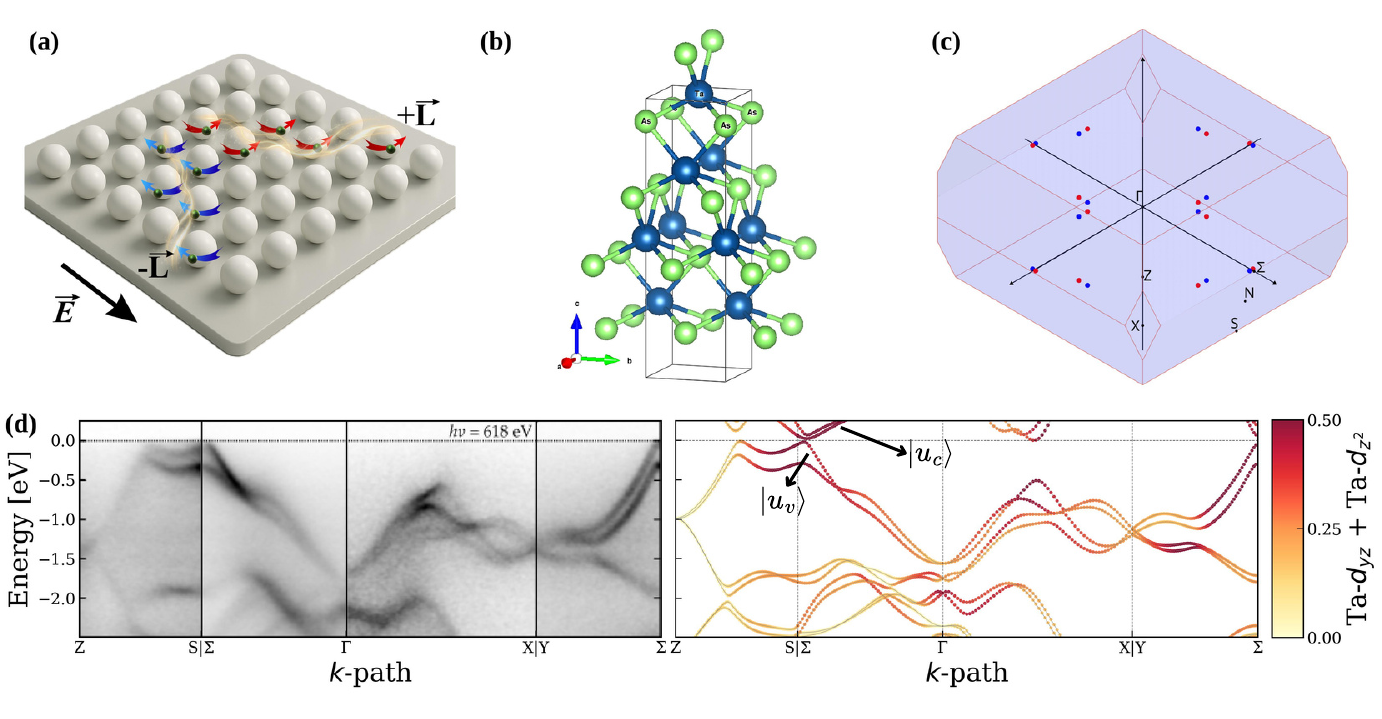}
\caption{a) Illustration of the orbital Hall effect, where an external electric field $\vec{E}$ produces a transverse, charge-neutral flow of orbital angular momenta. Opposite OAM-polarized electrons move in opposite directions. b) Conventional unit cell of TaAs. The TaAs family (TaAs, TaP, NbAs, NbP) crystallizes in space group $I4_1md$ (point group $C_{4v}$), that lacks inversion symmetry. The blue spheres represent the Ta/Nb atoms, the green ones the As/P atoms.  c) Brillouin zone of TaAs with Weyl nodes. This family of materials hosts 12 pairs of symmetry-robust Weyl nodes and associated surface Fermi arcs \cite{WSM_TaAs, TaAs2, TaAs3, TaAs4}. The presence of time-reversal symmetry prescribes that for each Weyl point at $\mathbf{k_0}$ there is a partner at $-\mathbf{k_0}$ with the same chirality (matching colours). d) \textit{Left}: ARPES band structure of TaAs along a high-symmetry path. \textit{Right}: DFT band structure along the same high-symmetry path. The bands are coloured based on the Ta-$(d_{z^2} + d_{yz})$ orbital weight, resulting in a good agreement with the ARPES intensity. The valence and conduction bands of the Weyl cone are highlighted by arrows.}
\label{fig1}
\end{figure*}

The notion that Bloch states can possess not only spin but also orbital angular momentum (OAM) \cite{Bernevig2005Orbitronics, Park2011OAMRashba, Sunko2017MaximalRashbaLike, Go2018OrbitalTexture, Choi2023OHEinTi, Ding2022OREMR, ElHamdi2023InverseORE, BurgosAtencia2024OAMReview} has recently attracted great attention in condensed matter research. This follows particularly from two key characteristics: first, OAM can be associated with Berry curvature \cite{Vanderbilt2018BerryPhases, Schueler2020}, making it an important physical observable for the investigation of topological quantum matter \cite{park_orbital-angular-momentum_2011,park_orbital_2012, park_chiral_2012, unzelmann_orbital-driven_2020-1, Unzelmann2021, brinkman_chirality-driven_2024, Figgemeier2025, oh2026observationspinfreeinteratomicorbital, oh2026pwaveorbitalangularmomentum}.
Second, the concept of OAM in condensed matter has led to the rise of \textit{Orbitronics}, an emergent field describing orbital analogs of spintronic phenomena \cite{BurgosAtencia2024OAMReview,go2021orbitronics,cysne2025orbitronics, Fukami2025}. Founded on the idea of harnessing intrinsic orbital currents for information storage and transport, it has inspired promising energy-efficient and low-dissipation technologies.
Indeed, orbital currents can be generated even in the absence of spin-orbit coupling (SOC), whereas SOC is the mechanism by which they may be subsequently converted into spin currents. For example, in the orbital Hall effect (OHE), a longitudinal electric field generates a transverse, charge-neutral flow of OAM (Fig.~\ref{fig1}a). This is, in turn, converted into a spin current in the presence of SOC, leading to the well-established spin Hall effect (SHE). While the OHE can thus be seen as a precursor to the SHE, conceptually the two phenomena only differ in the underlying degree of freedom transported by the transverse current: orbital angular momentum in the former and spin angular momentum in the latter.

Early theoretical work anticipated sizable orbital responses in transition metals~\cite{Tanaka2008PRB,Kontani2009PRL} and recent first-principles studies have revealed orbital Hall conductivities (OHC) reaching values as large as $\sim 8000$ ($\hbar/e$) $(\Omega\,\mathrm{cm})^{-1}$, significantly exceeding typical spin Hall conductivities (SHC)~\cite{salemi2022first, Go2024PRB, yang2023monopole}. Experimental signatures of the (inverse) OHE, have recently been provided by dynamical spin injection experiments on CoFeB/Cr/(Pt, Ta, W) multilayer structures~\cite{obinata2025experimental} and by Kerr microscopy measurements in the light metal Ti \cite{Choi2023OHEinTi}, where weak SOC allows to appreciate a genuine spin-free OAM physics.

In the context of spintronics, topological semimetals have been suggested to be promising materials with large SHCs. A seminal theory work (Ref.~\cite{Sun2016PRL}) predicted that the major contribution to the strong SHE in the paradigmatic class of TaAs Weyl semimetals originates from the topologically non-trivial \textit{hotspots} (i.e., the Weyl points) in the Brillouin zone.
In addition to that, experimental studies found that the characteristic Weyl band topology in TaAs is reflected into the intrinsic OAM texture, which emerges from the inversion symmetry breaking (ISB) in the non-centrosymmetric crystal structure ~\cite{ChulHee_Min_2019, Unzelmann2021, Figgemeier2025}.

These observations naturally suggest the investigation of orbitronics phenomena (especially the OHE) in non-centrosymmetric Weyl semimetals, which should enable one to address i) the interplay between non-trivial band topology and its associated, ISB-allowed OAM textures, ii) the predicted topology-driven SHE and iii) the inherent connection between the latter and the OHE. 
In a broader context, this will directly shed light on the microscopic mechanisms underlying OAM transport in topological quantum materials.

Motivated by this, we here present a detailed first-principles investigation of the orbital Hall effect in the TaAs family of Weyl semimetals.  The choice of transition metal monopnictides as a representative family of topological semimetals is --- beyond the aforementioned motivations --- particularly meaningful for a theoretical study because \textit{ab-initio} calculations generally match well with experimental results. This adds more significance to potential theoretical predictions. To highlight this, we refer to the striking match of our soft X-ray ARPES measurements and the calculated DFT bulk band structure, highlighted in Fig.~\ref{fig1}(d).

Using the atomic-center approximation (ACA) for the OAM operator inside the Kubo formula of the linear response theory, we find a strong orbital response that significantly exceeds its spin counterpart in the entire material family. Through systematic decomposition of the Kubo formula, we separate and quantify two relevant contributions to the OHE. The first ('diagonal') one is naturally associated with the transport of equilibrium OAM formed by ISB, while the second ('non-diagonal') term captures electric-field-induced interband ('non-diagonal') OAM processes ~\cite{Sahu2021,FARIDI2025170265}. The latter particularly includes the so-called orbital texture (OT) mechanism discussed in the context of centrosymmetric orbital Hall systems \cite{Go2018OrbitalTexture,Jo2018PRB, Fukami2025}.
Despite the strong intrinsic (diagonal) OAM texture in TaAs, our analysis surprisingly reveals a dominant role of the non-diagonal one. To elucidate the microscopic origin of this behavior, we develop an analytical Weyl model for the OHE. Based on adiabatic first-order perturbation theory, this model manifestly demonstrates that the non-diagonal (ND) channel, which captures the OT mechanism together with all possible band interference mechanisms, is preferentially enhanced in topological systems, especially if there are topological band-touching points.
Topology here emerges as a crucial factor in OAM physics: already at equilibrium, it shapes the highly non-trivial OAM textures observed in Refs.~\cite{ChulHee_Min_2019, Unzelmann2021, Figgemeier2025}, while out-of-equilibrium, it enters OAM transport enhancing it through inter-band hybridization~\cite{BernevigHughes+2013}.
Our study constitutes a further step towards a unified understanding of OAM transport, while at the same time designating the TaAs family as a promising platform for orbitronics applications.

All the electronic structure calculations were carried out within the framework of ab initio Density functional theory (DFT), implemented in the FPLO (Full-Potential Local-Orbital) package \cite{FPLO1,FPLO2}. The exchange–correlation energy was treated within the generalized gradient approximation (GGA), using the Perdew–Burke–Ernzerhof 96 (PBE) parameterization \cite{perdew1996generalized}. The spin-orbit interaction was implemented through fully relativistic calculations, within the built-in FPLO four-component Dirac formalism. The Brillouin zone was sampled using a Monkhorst–Pack \textbf{k}-point mesh of $12\times12\times12$. The relaxed structural parameters were taken from the Materials Project \cite{MP1,MP2,MP3}. 
The DFT orbitals of each system were then projected onto maximally localized Wannier functions (MLWF) \cite{MLWF}. The atomic orbitals chosen as trial functions were Ta-5d, Nb-4d, As-4p and P-3p, resulting in models with just 32 MLWFs as a basis set.
All the operators of interest in this study were therefore built in the more computationally convenient Wannier gauge, relying on a in-house implementation. It is worth noting that the Wannier functions also provide the necessary localized basis for a well-defined OAM operator.

We start our analysis by quantitatively comparing the orbital and spin transport properties across the TaAs material family.
Within the linear-response theory, the orbital and spin Hall conductivities are defined using the orbital/spin current densities
\begin{equation}
j_{\alpha}^{\gamma;\hat{L}/\hat{S}}=\sigma_{\alpha\beta}^{\gamma; \hat{L}/\hat{S}}\mathcal{E}_{\beta}, \label{linearity}
\end{equation}
where $j_{\alpha}^{\gamma;\hat{L}/\hat{S}}$ is the orbital/spin current density along the $\alpha$ direction, corresponding to a flow of electrons with orbital/spin polarization along $\gamma$, generated by an external electric field parallel to the $\beta$ direction. The orbital/spin Hall conductivity $\sigma_{\alpha\beta}^{\gamma; \hat{L}/\hat{S}}$ arises from the BZ integral of the corresponding Berry curvature, whose numerical adaptation reads

\begin{equation}\label{OHC_DISCRETE}
    \sigma_{\alpha \beta}^{\gamma, \hat{L}/\hat{S}}(\mu, T) = - \frac{e}{N_k V_{uc}} \sum_{n\textbf{k}}f_{n,\mu}(\textbf{k},T)\Omega_{n,\alpha \beta}^{\gamma, \hat{L}/\hat{S}}(\textbf{k}) \textit{,}
\end{equation}
with $f_{n,\mu}$ being the Fermi Dirac distribution. These coefficients are evaluated in the independent-particle approximation, using the DFT-derived Wannier Hamiltonian and neglecting exchange-correlation kernel corrections to the linear response~\cite{OnidaRMP}.
The orbital/spin-projected Berry curvature within the Kubo formalism takes the form

\begin{equation}\label{OBC}
    \Omega_{n,\alpha \beta}^{\gamma, \hat{L}/\hat{S}}(\textbf{k}) =
    2\hbar \sum_{m \neq n} \frac{\mathrm{Im}\left[\bra{u_{n\textbf{k}}} \hat{J}_{\alpha}^{\gamma,\hat{L}/\hat{S}}\ket{u_{m\textbf{k}}} \bra{u_{m\textbf{k}}} \hat{v}_{\beta}\ket{u_{n\textbf{k}}}  \right]}
    {\left(E_{n\textbf{k}}-E_{m\textbf{k}} -i\delta^+\right)^2} \textit{.}
\end{equation} 
$V_{uc}$ and $N_k$ in Eq. \ref{OHC_DISCRETE} are the volume of the unit cell in real space and the number of points in the $k$-mesh over which we integrate, respectively. To obtain converged Hall conductivities, this number must be of the order of $\sim 10^7$ and we set it to $200\cross200\cross200$ by relying onto the Wannier interpolation \cite{WannierInterpolation_SHC}. 
In Eq. \ref{OBC} we add an infinitesimal broadening $\delta^+$, set to $10$\,meV  in all our calculations, to ensure numerical stability in the occurrence of degeneracies. $\hat{v}_{\alpha}=\frac{1}{\hbar}\frac{\partial H(\textbf{k})}{\partial k_{\alpha}}$ is the velocity operator~\cite{Go2024PRB,k1l2-g57n} and $\hat{J}_{\alpha}^{\gamma;\hat{A}}=\frac{1}{2}\left(\hat{v}_{\alpha}\hat{A}_{\gamma} + \hat{A}_{\gamma}\hat{v}_{\alpha}\right)$ is the orbital/spin-current operator, where $\hat{A}$ is either $\hat{L}$ or $\hat{S}$. The $u_{n\textbf{k}}(\textbf{r})=e^{-i\textbf{k}\cdot \textbf{r}}\psi_{n\textbf{k}}(\textbf{r})$ are the cell-periodic part of the Bloch's eigenstates and $E_{n\textbf{k}}$ the Bloch's eigenvalues.
\begin{table}[b!]
\caption{OHC and SHC components at charge neutrality, $\mu=0$
(units: $10^{3}\times(\hbar/e)\cdot\Omega^{-1}\cdot\mathrm{cm}^{-1}$). The SHCs results exhibit the same qualitative order of magnitude and sign trend as found in Ref.~\cite{Sun2016PRL}.}
\label{Tab1}
\centering
\begin{tabular}{lcccccc}
\toprule
& \multicolumn{3}{c}{OHC} & \multicolumn{3}{c}{SHC} \\
\cmidrule(lr){2-4}\cmidrule(lr){5-7}
Material & $\sigma^{z;\hat{L}}_{xy}$ & $\sigma^{y;\hat{L}}_{zx}$ & $\sigma^{x;\hat{L}}_{yz}$ 
         & $\sigma^{z;\hat{S}}_{xy}$ & $\sigma^{y;\hat{S}}_{zx}$ & $\sigma^{x;\hat{S}}_{yz}$ \\
\midrule
TaAs & 2.57 & 2.01 & 2.21 & -0.757 & -0.337 & -0.361 \\
TaP  & 2.74 & 2.20 & 2.38 & -0.372 & -0.337 & -0.295 \\
NbAs & 2.49 & 1.82 & 2.33 & -0.266 & -0.294 & -0.212 \\
NbP  & 2.60 & 2.16 & 2.43 & -0.146 & -0.356 & -0.254 \\
\bottomrule
\end{tabular}
\end{table}

In Table \ref{Tab1} we report the symmetry-allowed~\cite{Seemann2015PRB} non-zero components of the orbital and spin Hall conductivities at charge neutrality and at $T=0$ K.
Interestingly, for all compounds, the orbital response is about one order of magnitude larger than its spin counterpart.
In particular, in TaAs, both orbital and spin transport are predicted to be large, making the material suitable for next-generation SOT-MRAM (Spin-Orbit Torque Magnetic Random Access Memory)~\cite{OT, OT2, OT3}. In contrast, Nb-based compounds with large OHC and small SHC, could be relevant for experimental efforts to disentangle orbital transport from spin transport~\cite{Choi2023OHEinTi}. \\
\begin{figure}[!t]
\centering
\includegraphics[width=\columnwidth,angle=0,clip=true]{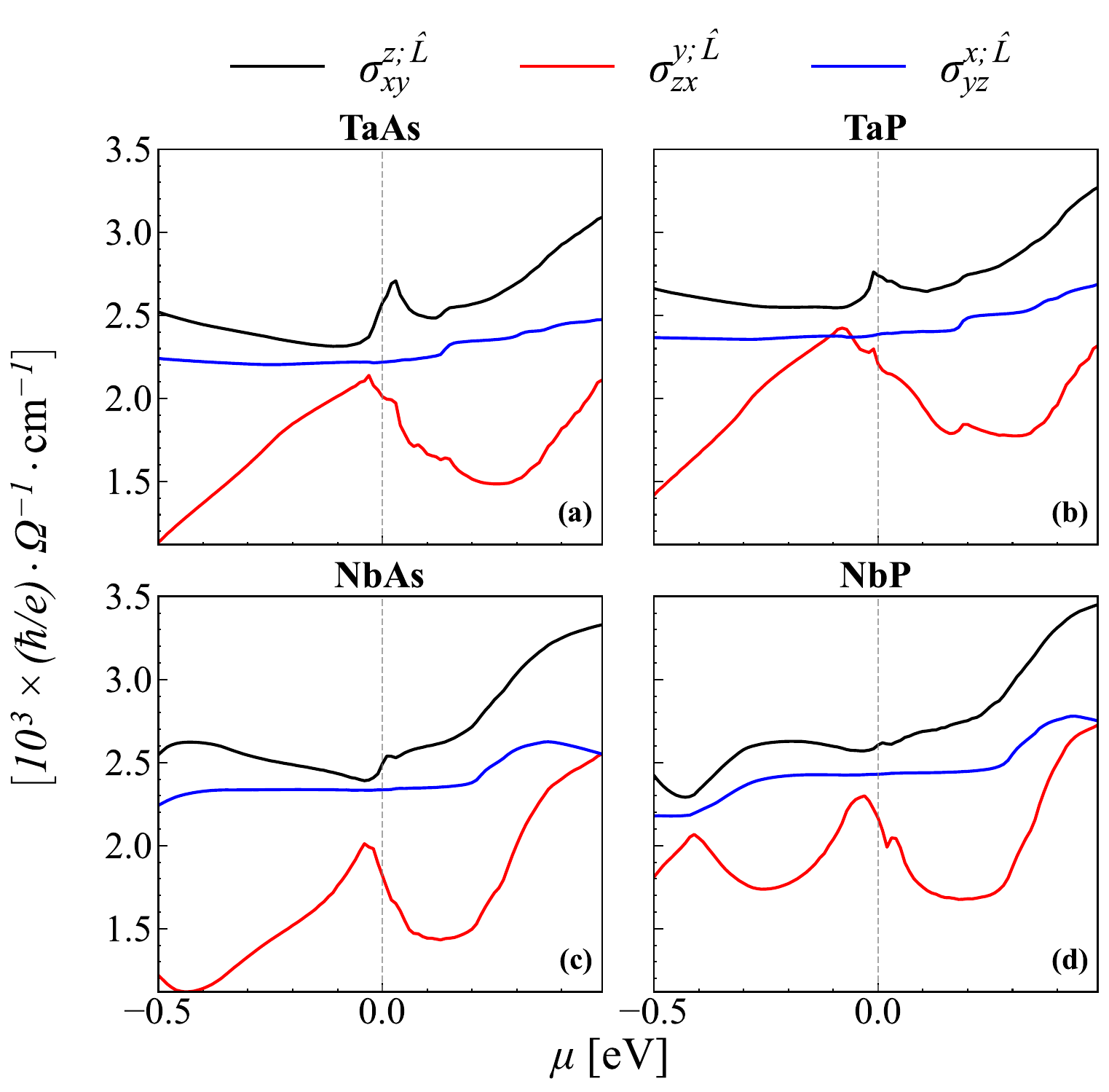}
\caption{a) Chemical potential dependence of the orbital Hall conductivity for the TaAs family of Weyl semimetals. The Fermi energy is set to zero at the charge neutral point.}
\label{Fig2}
\end{figure}
\indent Fig. \ref{Fig2} shows the chemical potential dependence of OHCs around the charge neutral point. This dependence was obtained within a rigid-band approximation. We highlight how the qualitative behaviors are similar between compounds based on the same transition metal. Further, we note that the overall magnitude of the OHE is independent of SOC (see Fig.~\ref{S-fig:no_soc_OHC_TaAs}~\cite{Supp}).
Spin-orbit interaction, however, can significantly influence its $\mu$-dependence. In particular, the behavior in Fig.~\ref{S-fig:no_soc_OHC_TaAs}~\cite{Supp} is quite similar to the one for Nb-based compounds (Fig. \ref{Fig2} (c) and (d)), which have a smaller atomic SOC strength.
This prominent qualitative dependence on SOC can be traced back to the Kubo formula (Eq.~\ref{OBC}) and its quadratic dependence on the inverse of the energies~\cite{Bernevig2005Orbitronics}.

\begin{figure}[!t]
\centering
\includegraphics[width=\columnwidth,angle=0,clip=true]{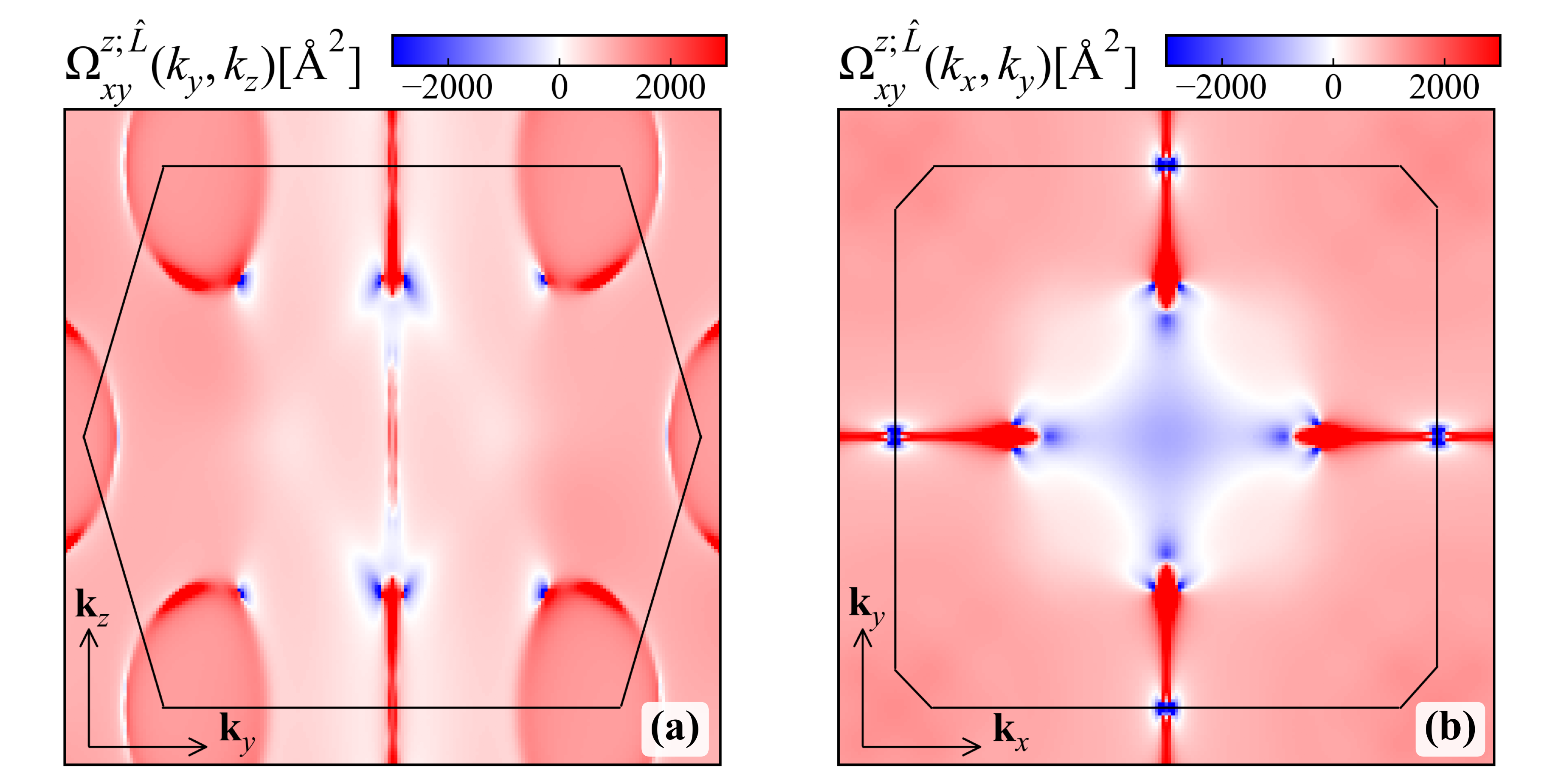}
\caption{$\mathbf{k}$-space texture of the orbital Berry curvature of TaAs projected on the a) $k_y-k_z$ plane and the b) $k_x-k_y$ plane. The projections were obtained by summing all the contributions along the a) $k_z$ and b) $k_x$ directions, respectively. The black contours indicate the first BZ boundaries.}
\label{Fig3}
\end{figure}

To dive deeper into the microscopic origin of orbital transport in Weyl semimetals, we now take TaAs as the representative and visualize its orbital Berry curvature (OBC) texture (Eq. \ref{OBC}) across the 3D bulk BZ. The OBC projected on the $k_z - k_y$ and $k_y - k_x$ planes are depicted in Fig.~\ref{Fig3} (a) and (b), respectively.
It appears that the highest contributions arise from the topologically "active" regions in $k$-space, i.e., the Weyl points (compare with the locations of the Weyl points in Fig.\ref{S-fig:planar_BZ}).
This immediately suggests that the characteristic Weyl topology may play a crucial role in the strength of the orbital transport in Weyl semimetals, in the same way as topology was predicted to be pivotal to SHE in TaAs \cite{Sun2016PRL}.

Based on this, the fact that the topological band geometry around the Weyl points in TaAs reflects itself in the inherent OAM texture \cite{Unzelmann2021} may suggest that the \textit{diagonal} contributions to the OHE play an important role. This naturally raises the question of how much the \textit{non-diagonal} contributions, which include for example the orbital texture mechanism (predicted to be a generic microscopic driving force for the OHE and SHE \cite{Go2018OrbitalTexture}) are significant relatively to the \textit{diagonal} ones.\\
To address this, let us first examine the respective contributions of the two mechanisms from a quantitative perspective.
We proceeded by unpacking Eq. \ref{OBC} for $\hat{A}=\hat{L}$ into two distinct parts: the \textit{diagonal} (D) $\Omega^{\gamma;D}$ and \textit{non-diagonal} (ND) $\Omega^{\gamma;ND}$ one, which,  due to the linearity regime of the response theory, simply sum up to give the total observable conductivity.  The D and ND attributes refer to the OAM operator, $\hat{L}$:

\begin{subequations}
\begin{align}
&\Omega_{n,\alpha \beta}^{\gamma; D}(\textbf{k}) = \hbar \Im \sum_{m \neq n} \left[ \frac{v_{\alpha}^{nm}v_{\beta}^{mn}}{(E_{n\textbf{k}}-E_{m\textbf{k}})^2} \times \big( L_{\gamma}^{mm} + L_{\gamma}^{nn} \big) \right] \label{OHC_NCS},\\
&\begin{aligned}
&\Omega_{n,\alpha \beta}^{\gamma; ND}(\textbf{k}) =  \hbar \Im \sum_{m \neq n} \Bigg\{ \frac{v_{\beta}^{mn}}{(E_{n\textbf{k}}-E_{m\textbf{k}})^2} \times \\ &\bigg[L_{\gamma}^{nm}(v_{\alpha}^{nn}+v_{\alpha}^{mm})  + \sum_{n' \neq n, m} \left( v_{\alpha}^{nn'} L_{\gamma}^{n'm} + L_{\gamma}^{nn'} v_{\alpha}^{n'm} \right) \bigg] \Bigg\}.
\end{aligned} \label{OHC_CS}
\end{align}
\end{subequations}

\noindent The detailed derivation can be found in the Supplementary Materials~\cite{Supp}.
Notice how in $\Omega^{\gamma;D}$ only the diagonal matrix elements $L_{\gamma}^{nn}=\langle u_{n\textbf{k}} | \hat{L} | u_{n\textbf{k}} \rangle$ appear. This contribution captures the transport of the \textit{equilibrium} OAM, the one that the \textbf{k}-states of the material intrinsically possesses, because the absence of an inversion point allows for it. On the other hand, $\Omega^{\gamma;ND}$ incorporates all the remaining, non-diagonal, matrix elements $L_{\gamma}^{nm}=\langle u_{n\textbf{k}} | \hat{L} | u_{m\textbf{k}} \rangle$, i.e. all the ones originated from inter-band hybridization mechanisms. 

\begin{figure}[!b]
\centering
\includegraphics[width=0.8\columnwidth,angle=0,clip=true]{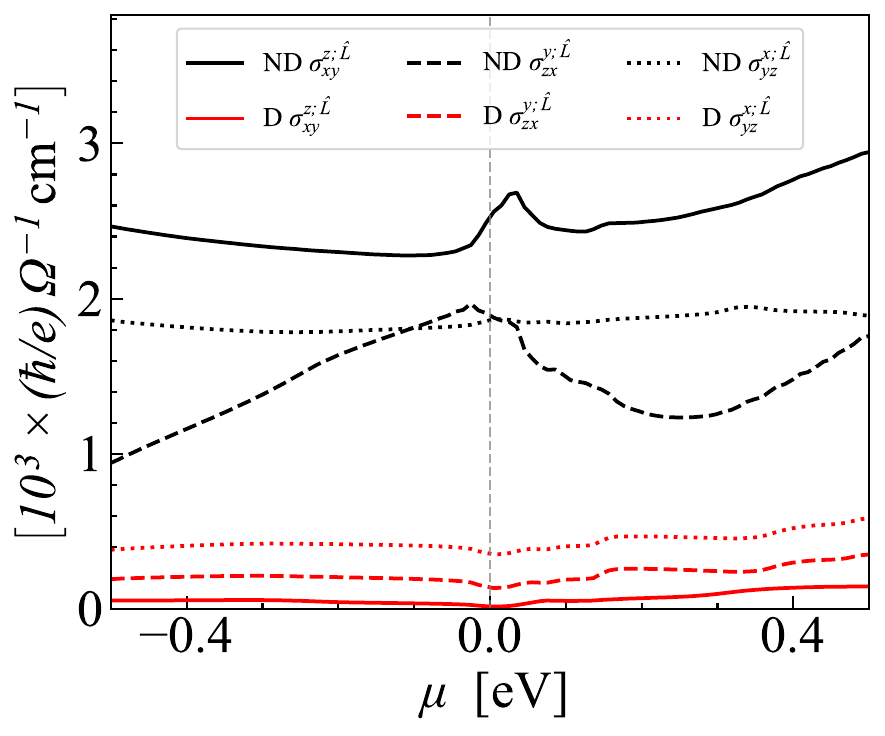}
\caption{Orbital Hall conductivity of TaAs as a function of Fermi energy, decomposed into diagonal and non-diagonal contributions. Red curves show the diagonal terms, while black curves show the non-diagonal terms for different orbital-current tensor components.}
\label{Fig4}
\end{figure}

In Fig. \ref{Fig4}, we plot the diagonal and non-diagonal contributions to orbital the Hall conductivity, calculated from the respective OBC $\Omega^{D}$ and $\Omega^{ND}$.
Surprisingly, the magnitudes of the non-diagonal components are strikingly dominant over the diagonal ones. This remains valid throughout the whole family of materials as well, as shown in Fig. \ref{S-FigS4}~\cite{Supp}.

To better understand these results, we analyze the
electric-field-induced orbital polarization within a minimal Weyl model $\hat{H_0}(\mathbf{k})=\hbar v\,\mathbf{k}\cdot\bm{\hat{\tau}}$~\cite{Supp}. Being $\rvert u_{v/c}(\mathbf{k})\rangle$ the valence and conduction Weyl eigenstates and $\mathcal{E}_{\alpha}$ the electric field component along $\alpha$, the dynamically-induced OAM for the valence band is
\begin{equation}
\delta\langle \hat{L}_{\gamma} \rangle_v
=2e \mathcal{E}_{\alpha}\,\mathrm{Im}\!\left[
\frac{L_{\gamma}^{v c}\,\langle u_{c} \rvert \partial_{k_{\alpha}} u_v \rangle}
{E_v-E_{c}}
\right],
\label{eq:Weyl_deltaL_appendix}
\end{equation}
where $L_{\gamma}^{vc}(\mathbf{k}) =: \langle u_{c}(\mathbf{k})\rvert \hat{L}_{\gamma}\rvert u_v(\mathbf{k})\rangle$ is the non-diagonal OAM matrix element. We recognize $\langle u_{c}(\mathbf{k})\rvert \partial_{k_x}u_v(\mathbf{k})\rangle$ as the Berry connection matrix element between the two Bloch's states, which geometrically describes their relative changes across the node.
Importantly, the Weyl dispersion gives this induced OAM the correct transverse parity. For \(E\parallel x\), the induced \(L_z\) contains a component odd in \(k_y\), so that its product with the transverse velocity \(v_y\), as entering \(J_y^{z,L}=\frac{1}{2}\{v_y,L_z\}\), is even under \(k_y\rightarrow -k_y\) (see Eq. \ref{S-parity}) and can contribute to the OHC.
Assuming a non-zero numerator, when the additional correction to $\langle \hat{L}_{\gamma} \rangle$ in Eq.~\ref{eq:Weyl_deltaL_appendix} is inserted into the response of the orbital-current operator, it gives rise to the non-diagonal contribution to the orbital Berry curvature (Eq. \ref{OBC}). The additional energy denominator associated with the field-induced hybridization combines with the denominator in the Kubo formula, making the ND channel particularly sensitive to small band splittings near band degeneracies. This provides the non-diagonal channel with a denominator of the form $(E_v - E_c)^{-n}$, with $n\geq 3$ \cite{footnote, Vanderbilt2018BerryPhases}, while the ISB channel always goes like $(E_v - E_c)^{-(n-1)}$. This enhanced sensitivity to the energy degeneracies makes the ND channel the favorite one when approaching band-touching points. 
However the crucial physics regarding the OHE is inside the numerator:  the ND channel is not active unless both the non-diagonal OAM and the Berry connection matrix elements are non-zero, which guarantees a finite \textit{gauge-invariant} orbital-geometric product \( \mathrm{Im}[L^{vc}_\gamma\langle u_c|\partial_{k_\alpha}u_v\rangle] \). 
Therefore, if the degeneracies also have a topological protection, the ND mechanisms are further enhanced by the Berry connection factor.
In our case, the two Weyl bands forming the cone are guaranteed by topology and symmetry to have an inherently non-zero and large Berry connection near the node~\cite{BernevigHughes+2013}. 

A non-vanishing $L_{\gamma}^{vc}(\mathbf{k})$, on the other hand, might be allowed and enhanced by many factors other than the orbital texture mechanism—such as the inter-site circulation of itinerant electrons, the spin-orbit-induced band mixing and lowered crystalline symmetries.
We now show evidence that, already at the simplest level of approximation, i.e. using the ACA for \(\hat L_z\) , the Weyl bands can have a finite $L_{z}^{vc}(\mathbf{k})$.
Since the major contribution to the low energy states comes from atomic $d$-orbitals (see Fig.~\ref{S-fig:fatbands}~\cite{Supp}), one can show that in the $d$-basis, the off-diagonal matrix elements of  \(\hat L_z\) are finite whenever, at each $\mathbf{k}$-point, the upper and lower bands contain \textit{different} orbital weights for those $d$-orbitals coupled by \(\hat L_z\), which are \(d_{xy}\leftrightarrow d_{x^2-y^2}\) or \(d_{yz}\leftrightarrow d_{xz}\)  (see Eq.~\ref{S-eq:ACA_multiorbital_codnition}~\cite{Supp}). 
In Fig.~\ref{fig1}(d), we show the bulk-sensitive soft X-ray ARPES measurement compared to the orbital-projected DFT bulk band structure calculation within a $\Gamma \Sigma \mathrm{X}$ equi-$k_z$ plane of the bulk BZ (Fig.~\ref{fig1}(c)). Overall, the ARPES data has a clear momentum dependence (indicating orbital texture), which matches well with the projected Ta $d_{yz}$ and $d_{z^2}$ orbital character. This orbital sensitivity can be explained by the circular light polarization in our experiment, which yields predominantly $E_y$ and $E_z$ light-electric-field components that are sensitive to these particular orbital characters (see SM~\cite{Supp}). 
The dominant orbital character changes from $d_{yz}$ in $u_{c}$ to $d_{z^2}$ in $u_{v}$ (see SM Fig.~\ref{S-fig:d_orbitals_grid}), which indicates a change of orbital texture across the Weyl bands.
However --- and this is the critical point here --- a pure $d_{yz}$ and $d_{z^2}$ character in the respective bands cannot cause the interband elements $\langle u_{c}(\textbf{k)} \rvert \hat{L}_z \lvert u_{v}(\textbf{k})\rangle$ to be non-zero. Here, the complex equilibrium OAM texture in the low-energy states, which also encodes the band topology in TaAs \cite{Unzelmann2021, Figgemeier2025}, comes into play. At equilibrium, inversion symmetry breaking leads to the formation of OAM, which is $L_y$ in the $M_{xz}$ mirror plane considered here. As such, the states have the form: $\rvert u_{v}(\mathbf{k})\rangle \propto \rvert d_{z^2} \rangle + \rvert d_{x^2-y^2} \rangle + i\,t_v(\mathbf{k}) \rvert d_{xz}\rangle$ and
$\rvert u_{c}(\mathbf{k})\rangle \propto \rvert d_{yz} \rangle + i\,t_c(\mathbf{k}) \rvert d_{xy}\rangle$, where $it_{v,c}(\mathbf{k})$ are complex ISB-scaled mixing terms. Thus, the ISB-induced multi-orbital character allows for interband couplings \(d_{xy}\leftrightarrow d_{x^2-y^2}\) and \(d_{yz}\leftrightarrow d_{xz}\) (see Eq. \ref{S-eq:ACA_multiorbital_codnition}) and thus non-zero $\langle u_{c}(\textbf{k)} \rvert \hat{L}_z \lvert u_{v}(\textbf{k})\rangle$.

The intuitive expectation that in a non-centrosymmetric crystal the diagonal contribution to the OHE is more important than the non-diagonal one has been proved not founded, in general. In our case, the intercession of the complex band geometry turns the result around, strongly enhancing the non-diagonal channel and promoting it to the main contributor of the OAM Hall transport.

Taken together, in the context of the OHE (and SHE), equilibrium and field-induced OAM are entangled by symmetry and the non-trivial topology in a complex manner. This picture combines two previously proposed perspectives, stating that the orbital and spin Hall effects are driven either by topology \cite{Sun2016PRL} or orbital texture \cite{Go2018OrbitalTexture}. As such, our study provides deep insights into the microscopic mechanisms underlying orbitronics (and spintronics) phenomena in topological materials. 
More broadly, the orbitronics framework naturally anticipates a large, SOC-robust orbital Hall effect in non-centrosymmetric Weyl semimetals. Our results for the TaAs family substantiate this expectation and reveal the microscopic origin of the effect in the synergy between band geometry, symmetry and interband interference. This understanding sets the stage for quantitative refinements and direct experimental tests through orbital torque measurements, optical orbital accumulation, and nonlocal transport~\cite{Lee2021NatCommun,Choi2023OHEinTi,Hayashi2023CommPhys}. Non-centrosymmetric Weyl semimetals thus emerge as a uniquely promising class of topological materials for next-generation orbitronics, where the interplay of topology, symmetry breaking, and orbital degrees of freedom can be harnessed by design.
{\it Acknowledgments.--}This work was supported by the European Research Council (ERC-2024-SyG-101167294; UnMySt), the Cluster of Excellence Advanced Imaging of Matter (AIM), Grupos Consolidados y Alto Rendimiento UPV/EHU, and Gobierno Vasco (IT1453-22). We acknowledge support from the Max Planck-New York City Center for Non-Equilibrium Quantum Phenomena. The Flatiron Institute is a division of the Simons Foundation. The authors gratefully acknowledge the Gauss Centre for Supercomputing e.V. (https://www.gauss-centre.eu) for funding this project by providing computing time on the GCS Supercomputer SuperMUC-NG at Leibniz Supercomputing Centre (https://www.lrz.de).
M\"U, TF, and FR thank the
Deutsche Forschungsgemeinschaft (DFG) supporting this through the Würzburg-Dresden Cluster of Excellence \textit{ctd.qmat} (EXC 2147, Project ID 390858490), SFB1170 'ToCoTronics' (projects A01 and B07), and Project RE1469-13-2 at JMU Würzburg.
The ASPHERE III photoemission spectroscopy instrument at beamline P04 was funded by the German Federal Ministry of Education and Research (BMBF) and the German Federal Ministry of Research, Technology, and Space (BMFTR) through the ErUM framework program (projects 05KS7WW1, 05K10WW2, 05K19WW2, and 05K25WW1 with the University of Würzburg)

\bibliography{biblio}

\clearpage
\onecolumngrid
\setcounter{figure}{0}
\setcounter{equation}{0}
\setcounter{section}{0}
\renewcommand{\thefigure}{S\arabic{figure}}
\renewcommand{\theequation}{S\arabic{equation}}
\renewcommand{\thesection}{S\Roman{section}}
\begin{center}
\textbf{\large Supplementary Material:\\ Orbital Hall Effect in Weyl Semimetals from Quantum Geometric Band Interference}
\end{center}
\vspace{1em}

\pagebreak

\setcounter{section}{0}
\setcounter{figure}{0}
\setcounter{equation}{0}
\setcounter{table}{0}
\tableofcontents
\clearpage

\subsection{Experimental methods: soft X-ray angle-resolved photoemission spectroscopy}

\subsubsection{Experimental setup}

\begin{figure}[h!]
    \centering
    \includegraphics[width=0.6\linewidth]{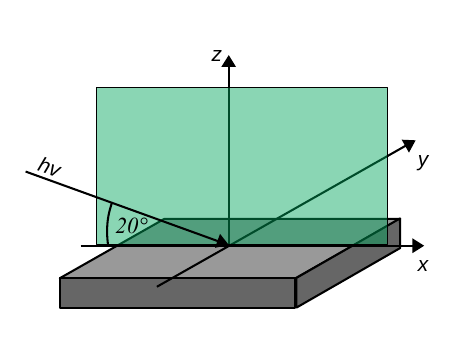}
    \caption{Experimental geometry of our soft x-ray ARPES setup. The $xz$ mirror plane (green) corresponds to the plane of light-incidence and detection plane of the emitted photoelectrons. That is, for each measured $k_\parallel$-path, the sample was oriented such that this path aligns with $xz$-plane.}
    \label{S-fig:exp_geometry}
\end{figure}

\subsection{Planar views of the Brillouin zone with the chirally charged Weyl points}

\begin{figure}[htbp]
    \centering
    \begin{subfigure}[b]{\textwidth}
        \centering
        \includegraphics[width=0.7\textwidth]{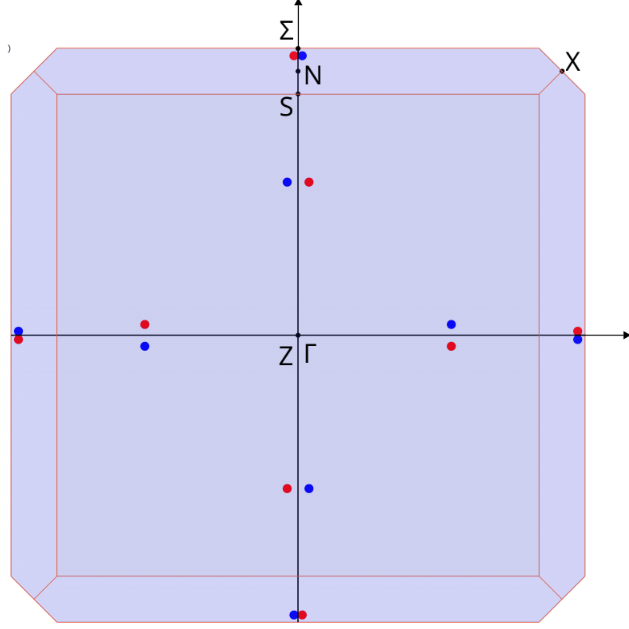}
        \label{S-fig:BZ_kz}
    \end{subfigure}
    
    \vspace{0.5cm}
    
    \begin{subfigure}[b]{\textwidth}
        \centering
        \includegraphics[width=0.7\textwidth]{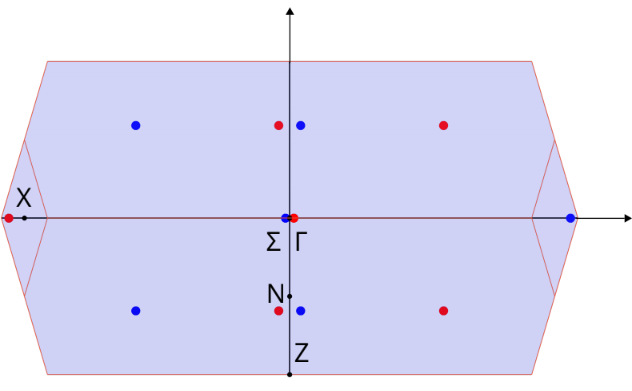}
        \label{S-fig:BZ_kx}
    \end{subfigure}
    
    \caption{Planar views of the TaAs BZ along (a) $k_z$ (b) $k_x$.}
    \label{S-fig:planar_BZ}
\end{figure}

\clearpage

\subsection{Tight-binding models for TaAs}

\begin{figure}[htbp]
    \centering
    \begin{subfigure}[b]{\textwidth}
        \centering
        \includegraphics[width=0.7\textwidth]{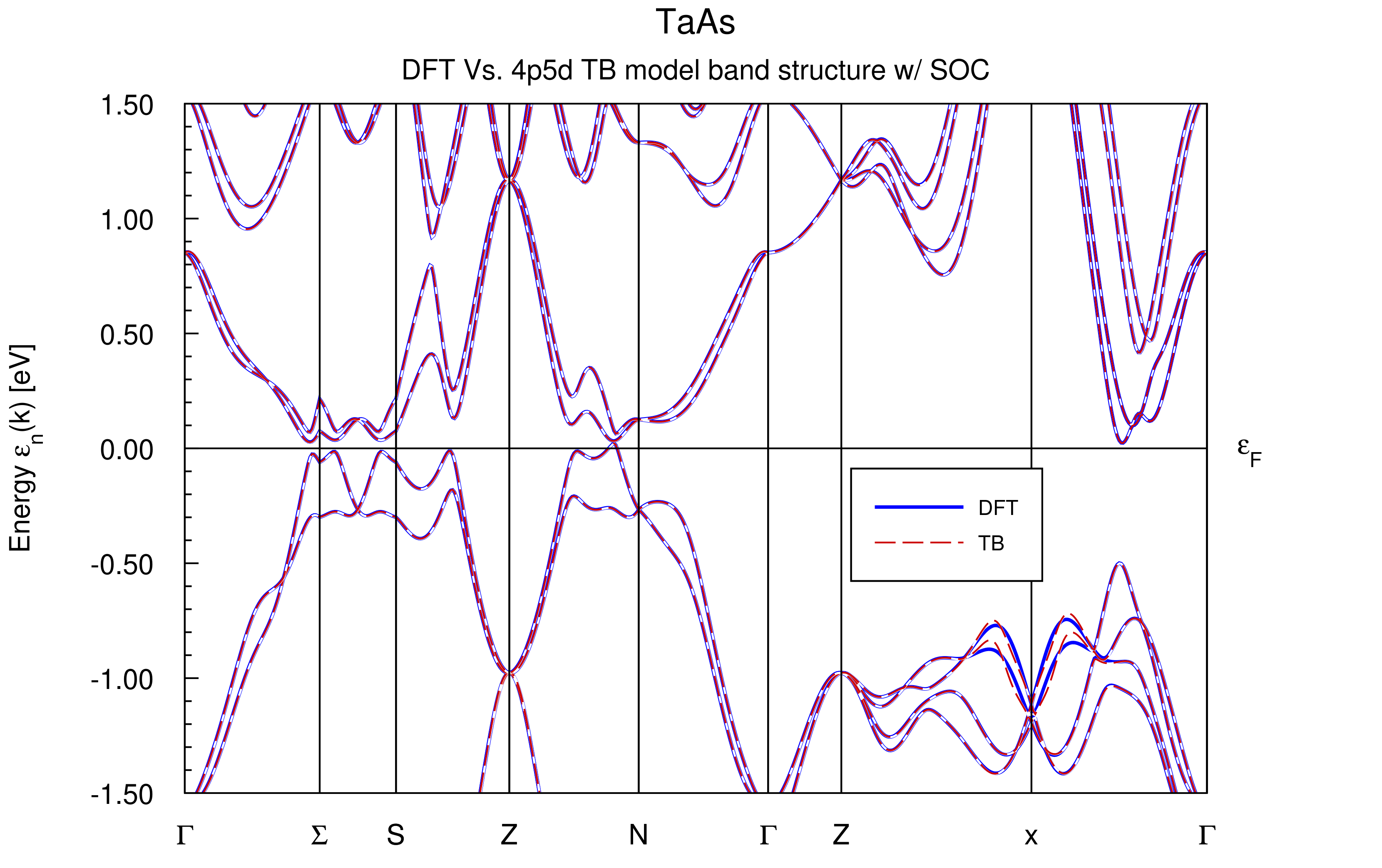}
        \label{S-fig:taas_soc}
    \end{subfigure}
    
    \vspace{0.5cm}
    
    \begin{subfigure}[b]{\textwidth}
        \centering
        \includegraphics[width=0.7\textwidth]{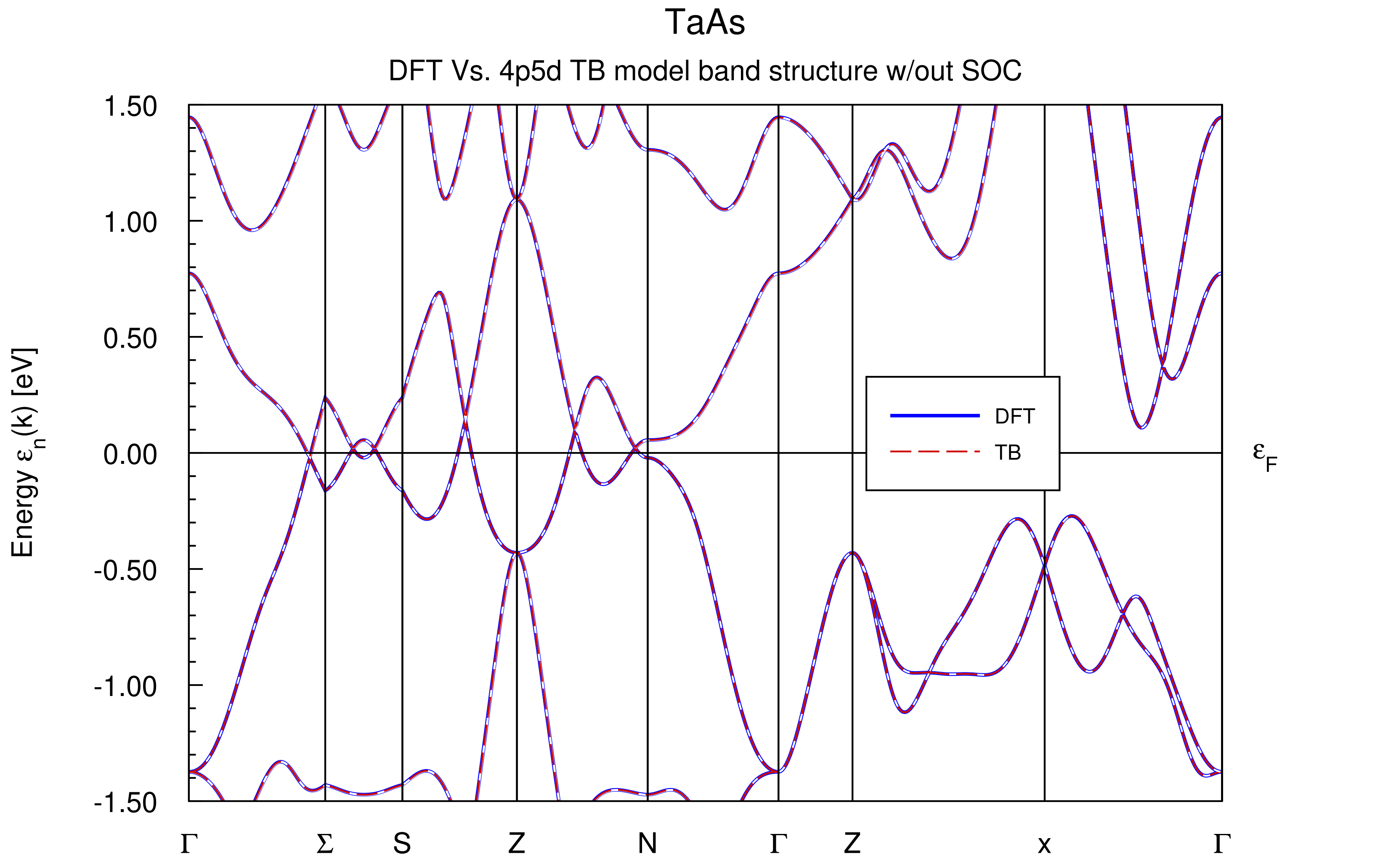}
        \label{S-fig:taas_zoom}
    \end{subfigure}
    
    \caption{Comparison of TaAs band structures. (a) with and (b) without SOC. For both cases there are the TB bands (Ta-5d, As-4p basis) on top of the DFT ones.}
    \label{S-fig:taas_comparison}
\end{figure}

\clearpage

\subsection{SOC-free orbital Hall conductivity in TaAs}

\begin{figure}[h!]
    \centering
    \includegraphics[width=0.6\linewidth]{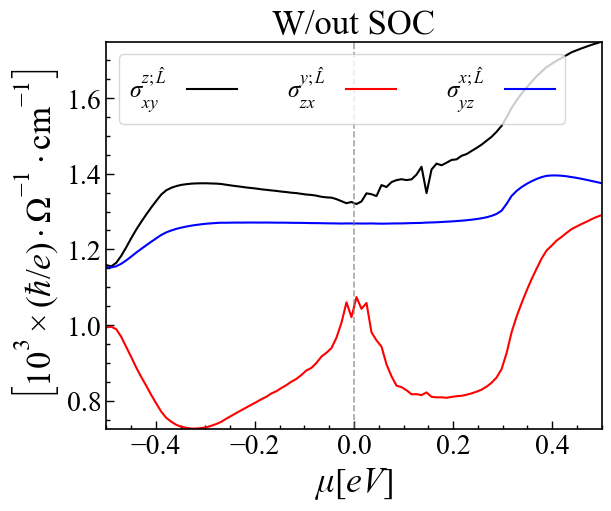}
    \caption{Chemical potential dependence of TaAs without the Spin-orbit interaction. The OHE is confirmed as a SOC-robust phenomenon, since the order of magnitude of the conductivity components remains unchanged. The qualitative behavior over the inspected range of $\mu$, if compared to the results in Fig. 2, is much more similar to the one found for the Nb-based compounds, where the SOC coupling is less relevant, being the Nb atom lighter than the Ta one.}
    \label{S-fig:no_soc_OHC_TaAs}
\end{figure}

\clearpage

\clearpage

\subsection{Orbital characters textures}

\begin{figure}[h!]
    \centering
    \includegraphics[width=\linewidth]{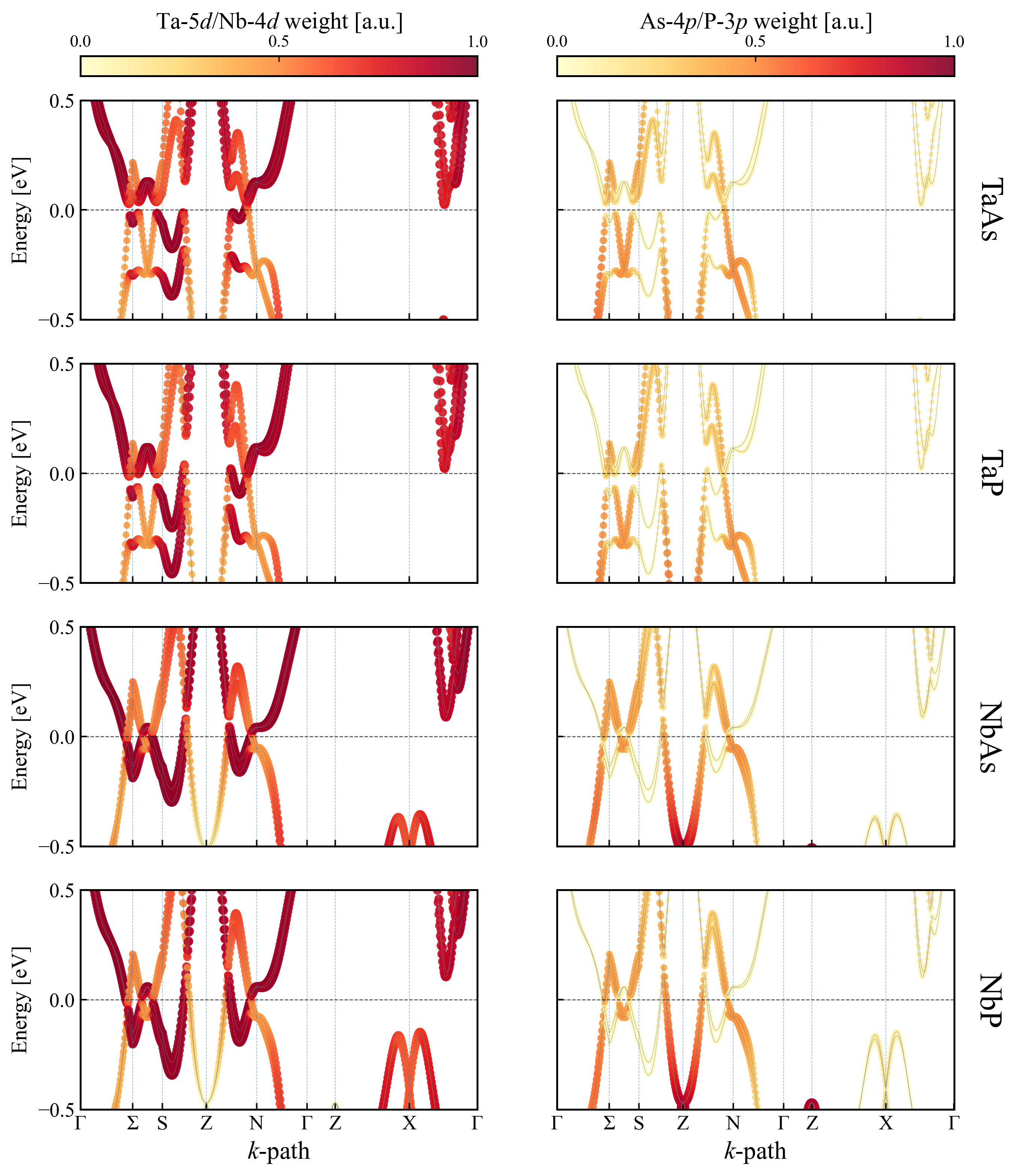}
    \caption{Comparison of the $d$ and $p$ orbital textures in the low-energy bands across our family of materials. SOC is included. }
    \label{S-fig:fatbands}
\end{figure}

\begin{figure}[p]
    \centering
 
    \begin{subfigure}[b]{0.49\textwidth}
        \includegraphics[width=\linewidth]{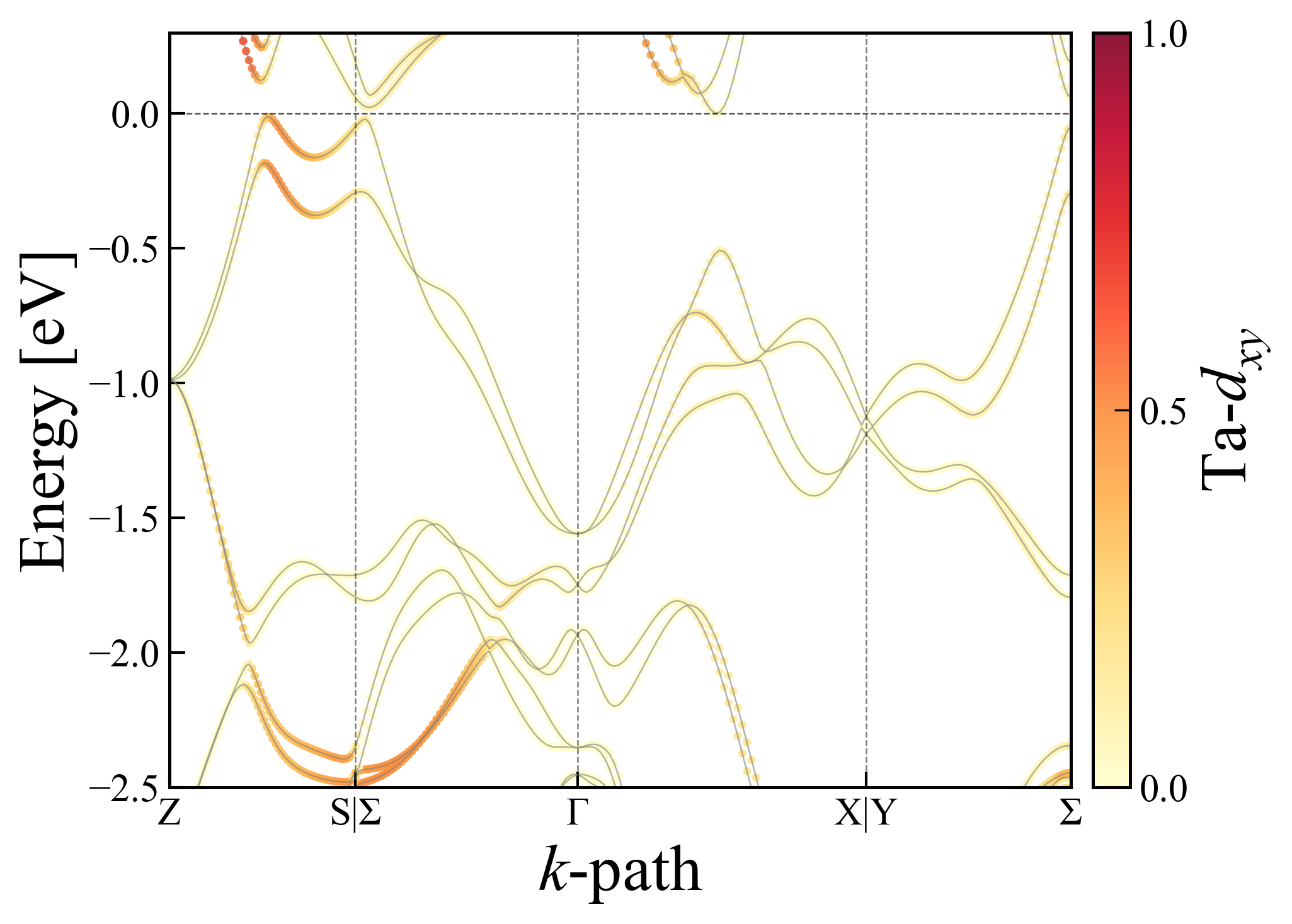}
        \label{S-fig:dxy}
    \end{subfigure}
    \begin{subfigure}[b]{0.49\textwidth}
        \includegraphics[width=\linewidth]{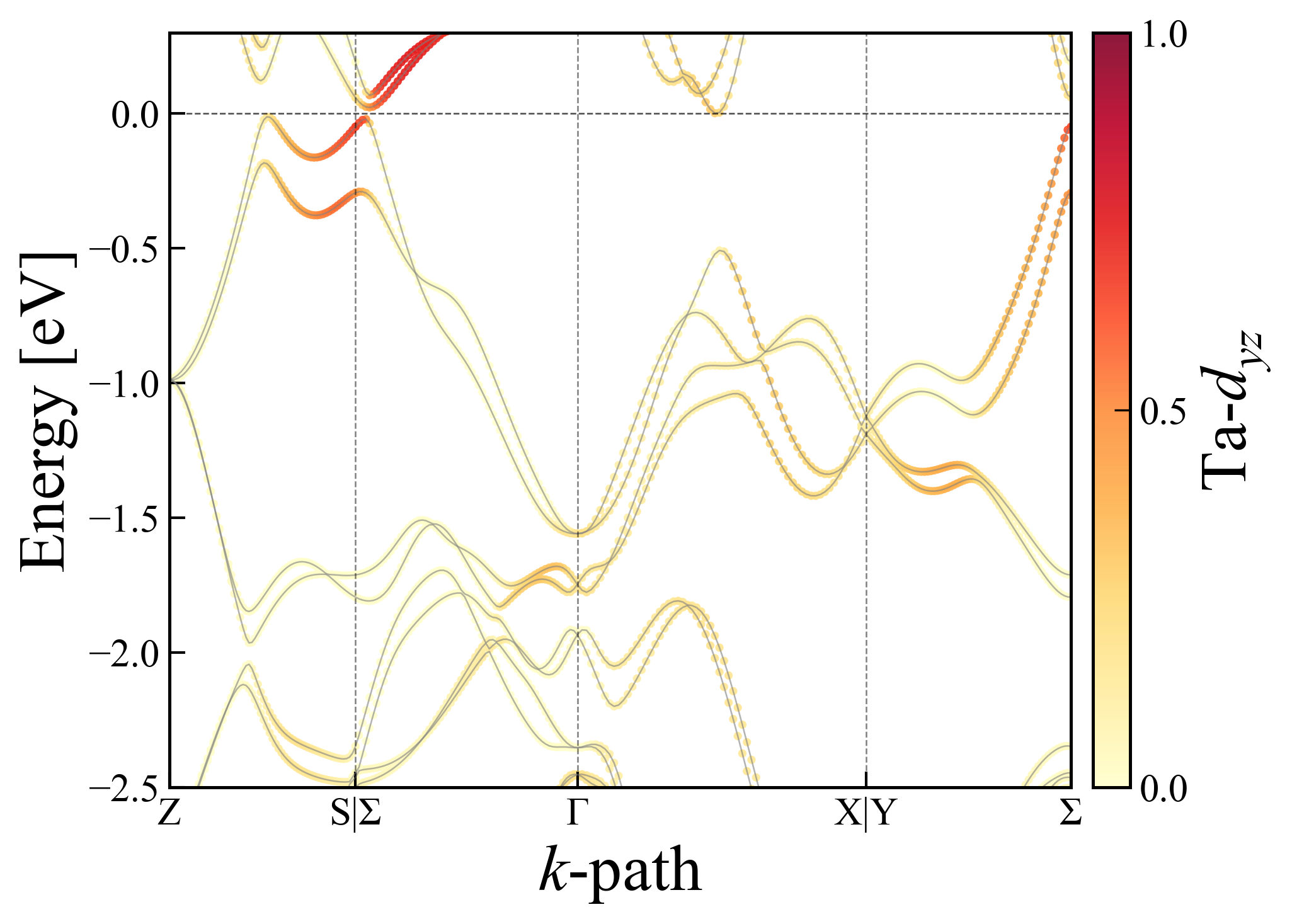}
        \label{S-fig:dyz}
    \end{subfigure}
    
    \begin{subfigure}[b]{0.49\textwidth}
        \includegraphics[width=\linewidth]{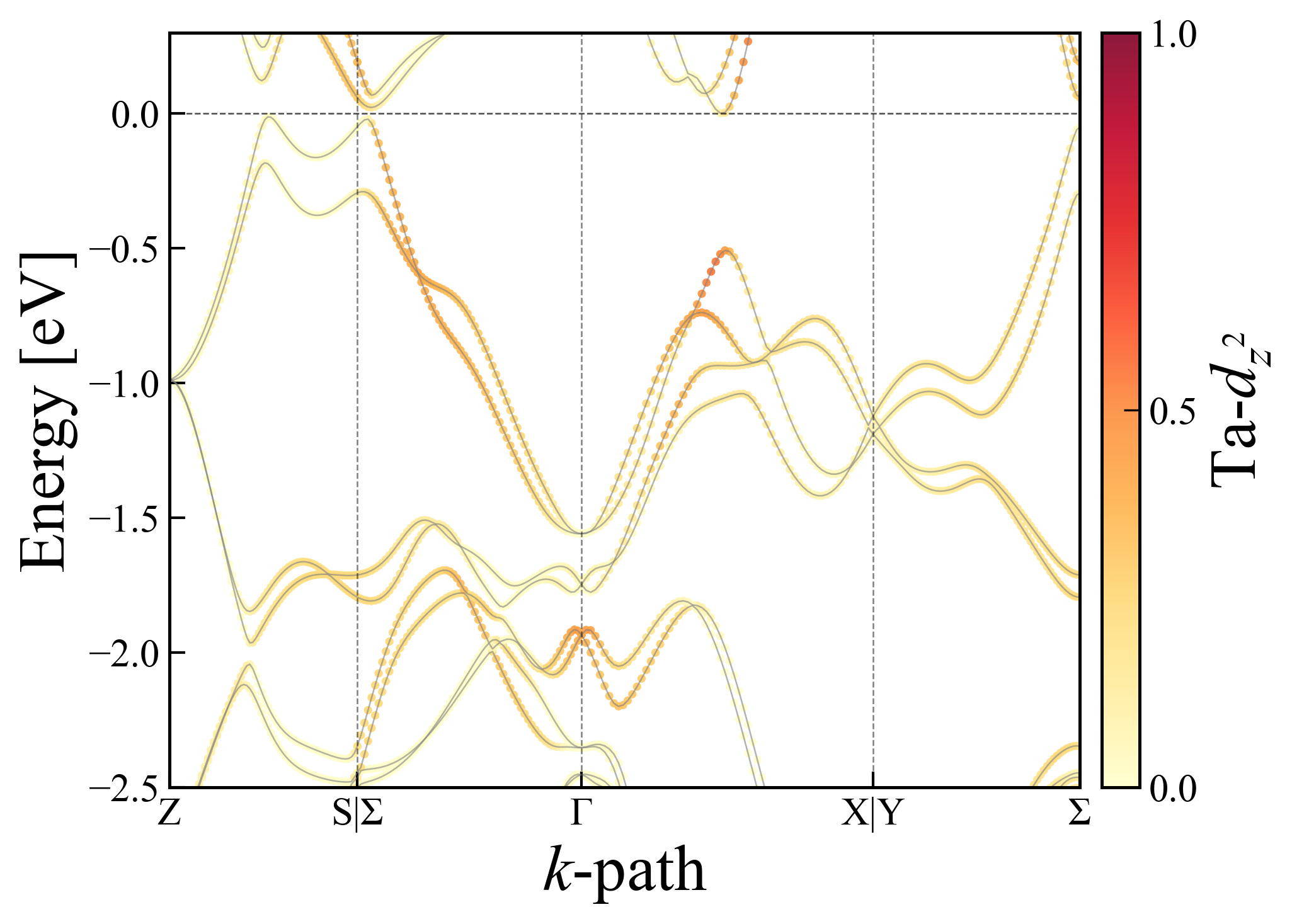}
        \label{S-fig:dz2}
    \end{subfigure}
 
    \begin{subfigure}[b]{0.49\textwidth}
        \includegraphics[width=\linewidth]{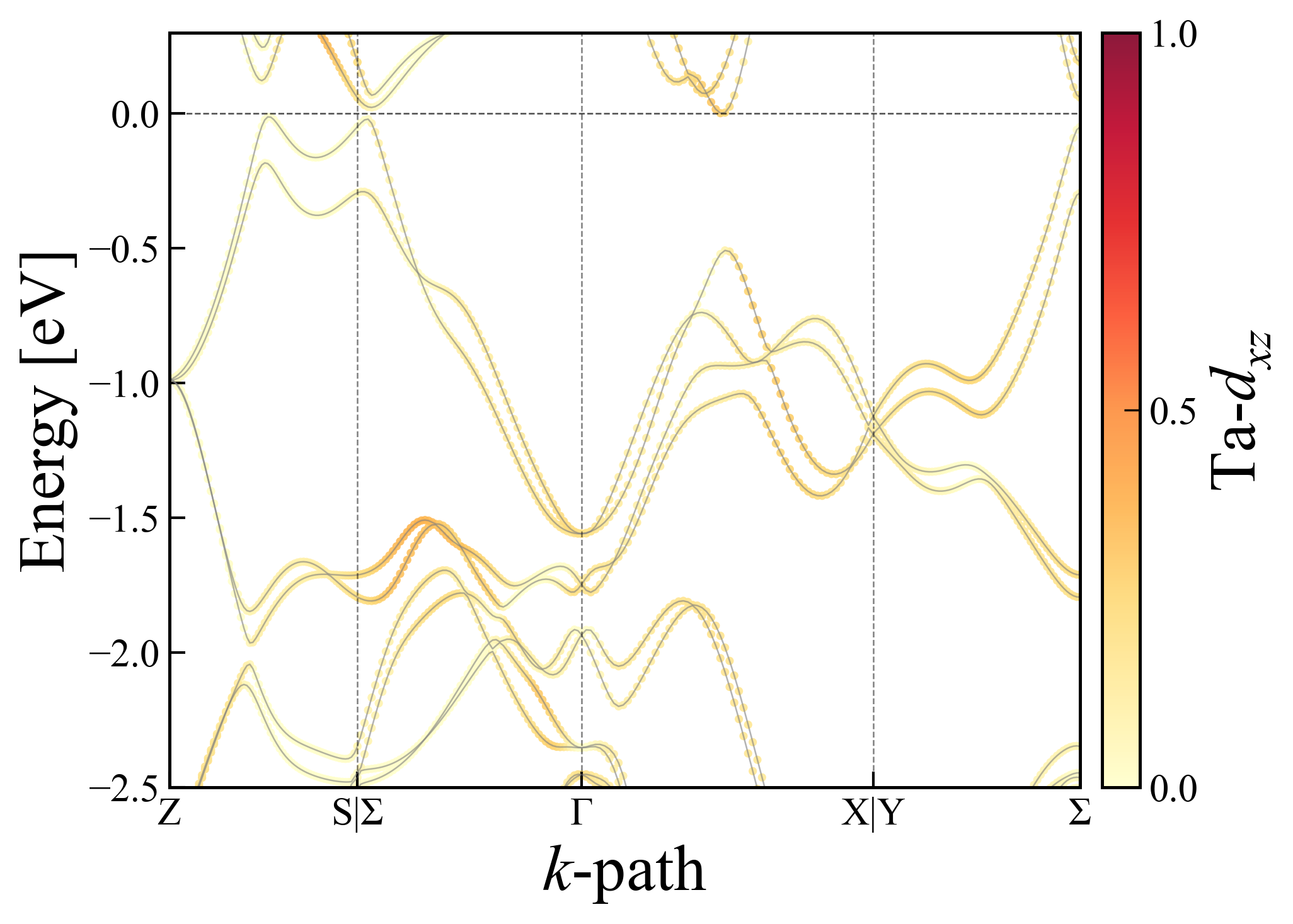}
        \label{S-fig:dxz}
    \end{subfigure}
    \begin{subfigure}[b]{0.49\textwidth}
        \includegraphics[width=\linewidth]{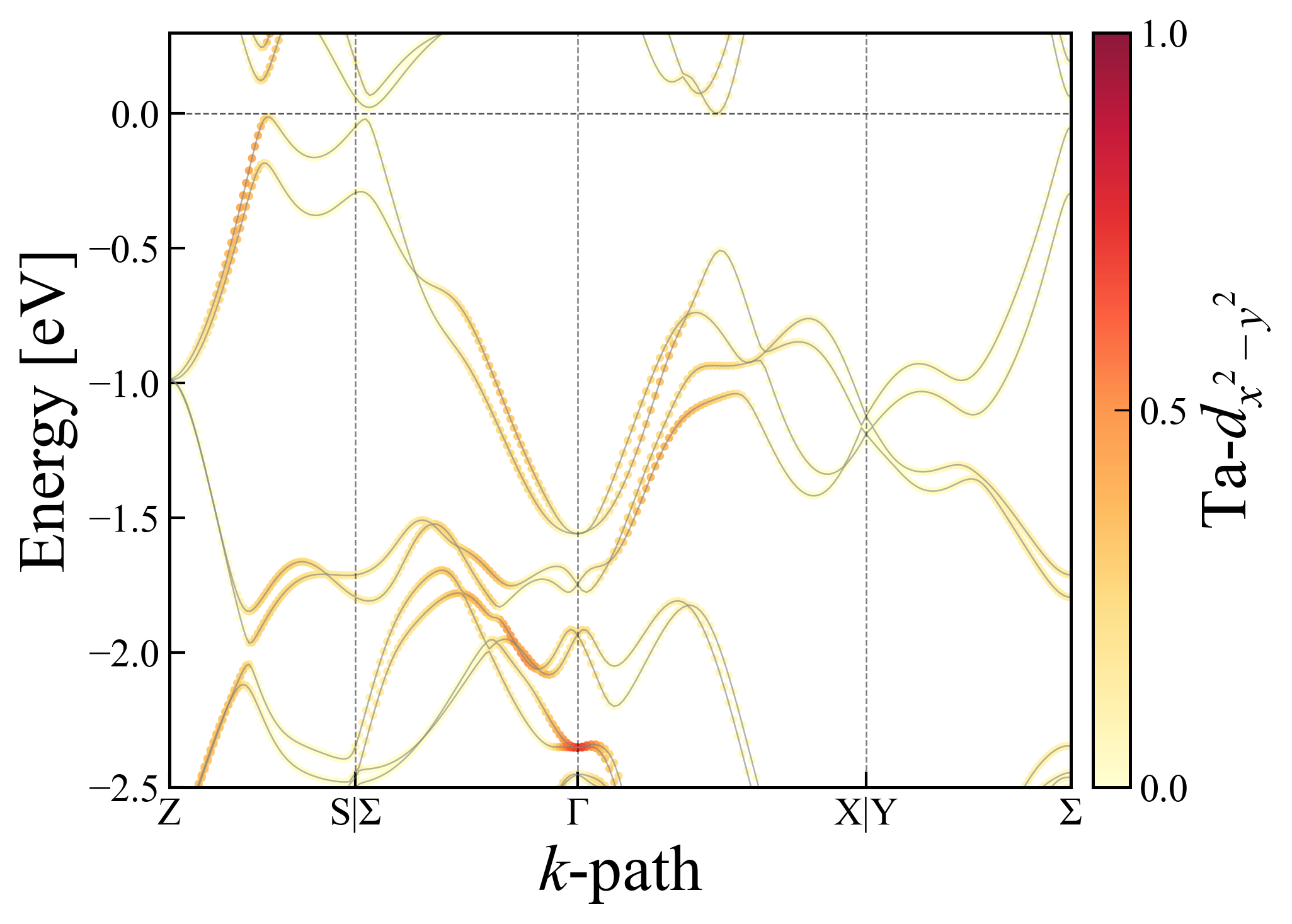}
        \label{S-fig:dx2y2}
    \end{subfigure}
 
    \caption{Projected TaAs band structure for each individual $d$-orbital character.}
    \label{S-fig:d_orbitals_grid}
\end{figure}

\clearpage

\subsection{OHC "D" and "ND" contributions for all materials}

\begin{figure}[h!]
\centering
\includegraphics[width=1\linewidth,angle=0,clip=true]{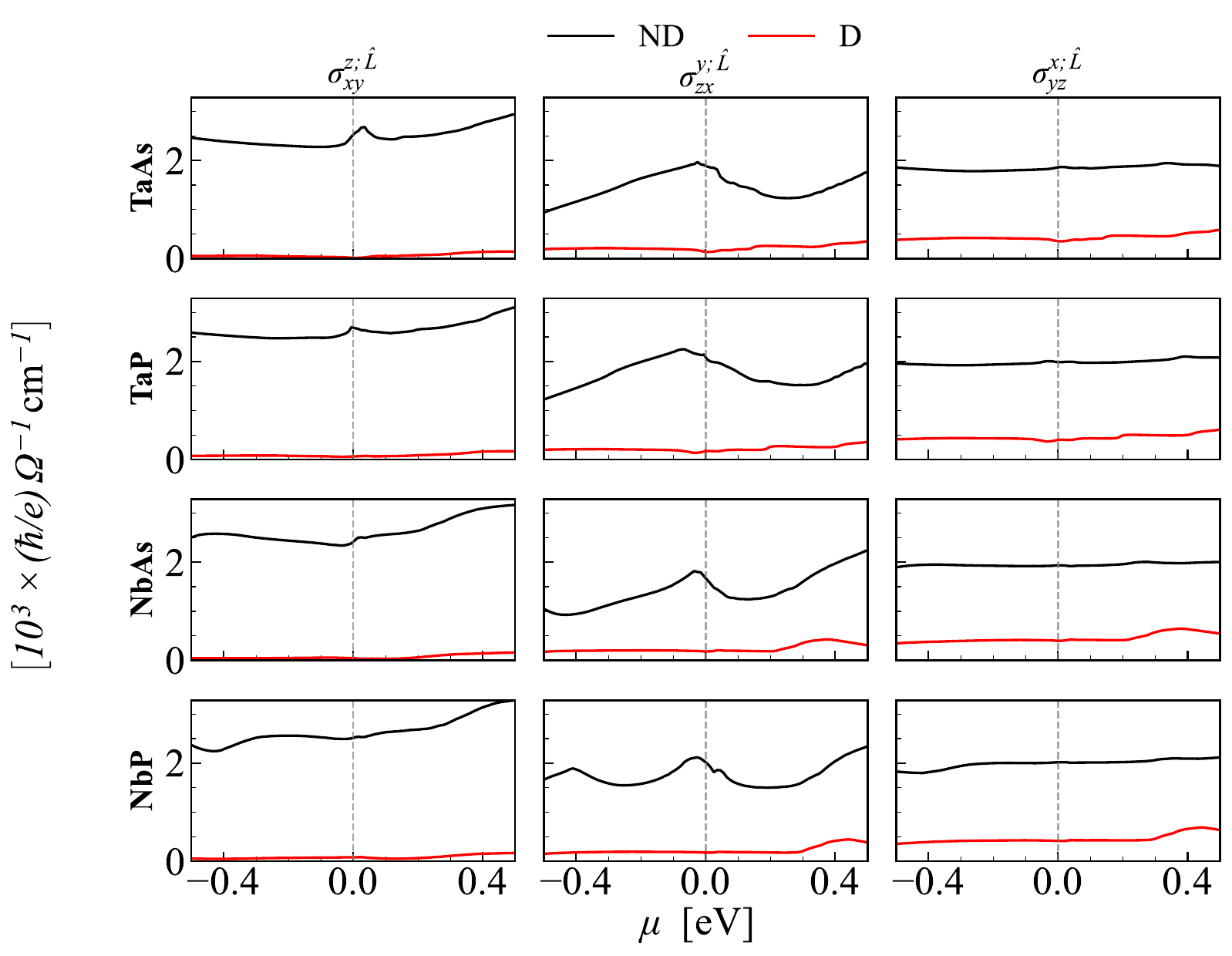}
\caption{Orbital Hall conductivity non-zero components unpacked in the \textit{diagonal} (D) and \textit{non-diagonal} (ND) contributions.}
\label{S-FigS4}
\end{figure}

\clearpage

\subsection{Detailed derivation for the decomposition of the Orbital Hall conductivity}\label{S-CS_NCS_derivation}

Within the linear response theory, the orbital Hall conductivity $\sigma_{\alpha\beta}^{\gamma; \hat{L}}$ is defined as the Brillouin zone integral of the orbital Berry curvature:

\begin{equation}\label{S-OHC_DISCRETE}
   \sigma_{\alpha \beta}^{\gamma, \hat{L}}(\mu, T)= -e\int_{BZ}\frac{d^3k}{(2\pi)^3}\sum_{n}f_{n,\mu}(\textbf{k},T)\Omega_{n,\alpha \beta}^{\gamma; \hat{L}}(\textbf{k})
\end{equation}
Where the orbitally-projected Berry curvature, according to the Kubo formalism, reads

\begin{equation}\label{S-OBC}
    \Omega_{n,\alpha \beta}^{\gamma, \hat{L}}(\textbf{k}) =
    2\hbar \sum_{m \neq n} \frac{\mathrm{Im}\left[\bra{u_{n\textbf{k}}} \hat{J}_{\alpha}^{\gamma,\hat{L}}\ket{u_{m\textbf{k}}} \bra{u_{m\textbf{k}}} \hat{v}_{\beta}\ket{u_{n\textbf{k}}}  \right]}
    {\left(E_{n\textbf{k}}-E_{m\textbf{k}} -i\delta^+\right)^2} \textit{.}
\end{equation}
With
\begin{equation}\label{S-orbital current op}
    \hat{J}_{\alpha}^{\gamma;\hat{L}}=\frac{1}{2}\{\hat{v}_{\alpha}, \hat{L}_{\gamma} \} = \frac{1}{2}\left(\hat{v}_{\alpha}\hat{L}_{\gamma} + \hat{L}_{\gamma}\hat{v}_{\alpha}\right) \textit{,}
\end{equation}the orbital-current operator. 

To investigate the microscopic mechanisms at the heart of the orbital Hall transport, we need to isolate the "pieces" of Eq. \ref{S-OBC} where the \textit{diagonal} (D) matrix elements of the OAM $L_{\gamma}^{nn}$ appear, from the ones with the \textit{non-diagonal} (ND) matrix elements, $L_{\gamma}^{nm}$. Then, exploiting the linearity of Eq. \ref{S-OHC_DISCRETE}, one immediately obtains the corresponding contributions to the conductivity.

We start by dissecting the matrix elements $\bra{u_{n\textbf{k}}}\left(\hat{v}_{\alpha}\hat{L}_{\gamma} + \hat{L}_{\gamma}\hat{v}_{\alpha}\right)\ket{u_{m\textbf{k}}}=\left(\hat{v}_{\alpha}\hat{L}_{\gamma} + \hat{L}_{\gamma}\hat{v}_{\alpha}\right)^{nm}$, with $m\neq n$, as prescribed by Eq. \ref{S-OBC}.
Let us notice that, obviously, $\left(\hat{v}_{\alpha}\hat{L}_{\gamma} + \hat{L}_{\gamma}\hat{v}_{\alpha}\right)^{nm}=\left(\hat{v}_{\alpha}\hat{L}_{\gamma}\right)^{nm} + \left(\hat{L}_{\gamma}\hat{v}_{\alpha}\right)^{nm}$. We then need to evaluate the elements of a product of operators.
For simplicity, let's consider two generic $N\cross N$ matrix operators $\mathcal{A}$ and $\mathcal{B}$:
\begin{align}
    \mathcal{A}=\begin{pmatrix}     
                A_{11} & A_{12} & ... & A_{1N} \\
                A_{21} & ... &   &  \\
                   ... &   & ... &  \\
                A_{N1} & A_{N2} & ... & A_{NN}
                \end{pmatrix} & \qquad
                 \mathcal{B}=\begin{pmatrix}
B_{11} & B_{12} & ... & B_{1N} \\
B_{21} & ... &   &  \\
... &   & ... &  \\
B_{N1} & B_{N2} & ... & B_{NN}.
\end{pmatrix}
\end{align}
The matrix elements $(\mathcal{A}\mathcal{B})^{nm}$ and $(\mathcal{B}\mathcal{A})^{nm}$ will then correspond to the scalar product between the $n-th$ row of the first operator and the $m-th$ column of the second one.
By explicit calculation we have, choosing $m>n$:
\begin{equation}\label{S-1.58}
\begin{split}
&(\mathcal{A}\mathcal{B})^{nm}=\begin{pmatrix}
A_{n1} & A_{n2} & ... & A_{nn} & ... & A_{nN}\end{pmatrix}\begin{pmatrix}
B_{1m} \\
B_{2m} \\
... \\
B_{mm} \\
... \\
B_{Nm}
\end{pmatrix} = \\ &= A_{n1}B_{1m}+A_{n2}B_{2m}+...+A_{nn}B_{nm}+...+A_{nm}B_{mm}+...+A_{nN}B_{Nm}.
\end{split}
\end{equation}

\begin{equation}\label{S-1.59}
\begin{split}
&(\mathcal{B}\mathcal{A})^{nm}=\begin{pmatrix}
B_{n1} & B_{n2} & ... & B_{nn} & ... & B_{nN}\end{pmatrix}\begin{pmatrix}
A_{1m} \\
A_{2m} \\
... \\
A_{mm} \\
... \\
A_{Nm} 
\end{pmatrix} = \\ &= B_{n1}A_{1m}+B_{n2}A_{2m}+...+B_{nn}A_{nm}+...+B_{nm}A_{mm}+...+B_{nN}A_{Nm}.
\end{split}   
\end{equation}
Adapting Eqs. \ref{S-1.58} and \ref{S-1.59} to our case, we shall now recast the orbital-current matrix elements by highlighting the terms proportional to at least one diagonal element of either $\hat{L}_{\gamma}$ or $\hat{v}_{\alpha}$:
\begin{equation}\label{S-1.60}
\begin{split}
\bra{u_{n\textbf{k}}}\hat{J}_{\alpha}^{\gamma;\hat{L}}\ket{u_{m\textbf{k}}}=\frac{1}{2}\left(  v_{\alpha}^{nn}L_{\gamma}^{nm} +  v_{\alpha}^{nm}L_{\gamma}^{mm} +\sum_{n^{\prime}\neq n, m} v_{\alpha}^{nn^{\prime}}L_{\gamma}^{n^{\prime}m}\right)+\\+\frac{1}{2}\left(  L_{\gamma}^{nn}v_{\alpha}^{nm} +  L_{\gamma}^{nm}v_{\alpha}^{mm} +\sum_{n^{\prime}\neq n, m} L_{\gamma}^{nn^{\prime}}v_{\alpha}^{n^{\prime}m}\right).
\end{split}    
\end{equation}
We insert Eq. \ref{S-1.60} into Eq. \ref{S-OBC}. In this passage, we shall now focus on the terms proportional only to the diagonal elements $L_{\gamma}^{nn}$ and isolate them. Notice how those are the terms directly proportional to the \textit{intrinsic} orbital magnetic moments of the $\ket{u_{n\textbf{k}}}$ states\footnote{All matrix elements above are \textbf{k}-dependent.}:
\begin{equation}    M_{\gamma}^n(\textbf{k})=\frac{g\mu_BL_{\gamma}^{nn}}{\hbar},
\end{equation}
where \textit{g} is the g-factor and $\mu_B = e\hbar/(2m)$ is the Bohr magneton.
These magnetic moments are present already at equilibrium in non-centrosymmetric crystals and by isolating them, we simultaneously quantify also all the remaining terms, which are necessarily due to the inter-band hybridization mechanism described in \cite{Go2018OrbitalTexture}. So we can write:

\begin{equation}
\begin{split}
\Omega_{n,\alpha \beta}^{\gamma; \hat{L}}(\textbf{k}) = \hbar \sum_{n \neq m} \frac{1}{(E_{n\textbf{k}}-E_{m\textbf{k}})^2} \times \Im \bigg[ v_{\alpha}^{nm} L_{\gamma}^{mm} v_{\beta}^{mn} + L_{\gamma}^{nn} v_{\alpha}^{nm} v_{\beta}^{mn} + \\
 v_{\alpha}^{nn} L_{\gamma}^{nm} v_{\beta}^{mn} + L_{\gamma}^{nm} v_{\alpha}^{mm} v_{\beta}^{mn} + 
 \sum_{n' \neq n, m} \left( v_{\alpha}^{nn'} L_{\gamma}^{n'm} + L_{\gamma}^{nn'} v_{\alpha}^{n'm} \right) v_{\beta}^{mn} \bigg].
\end{split}
\end{equation}
Now we can easily separate the two contributions:
\begin{subequations}
\begin{align}
&\Omega_{n,\alpha \beta}^{\gamma; D}(\textbf{k}) = \hbar \Im \sum_{m \neq n} \left[ \frac{v_{\alpha}^{nm}v_{\beta}^{mn}}{(E_{n\textbf{k}}-E_{m\textbf{k}})^2} \times \big( L_{\gamma}^{mm} + L_{\gamma}^{nn} \big) \right] \label{S-OHC_NCS},\\[5pt]
&\begin{aligned}
\Omega_{n,\alpha \beta}^{\gamma; ND}(\textbf{k}) = & \hbar \Im \sum_{m \neq n} \Bigg\{ \frac{v_{\beta}^{mn}}{(E_{n\textbf{k}}-E_{m\textbf{k}})^2} \times \bigg[L_{\gamma}^{nm}(v_{\alpha}^{nn}+v_{\alpha}^{mm}) + \\ & + \sum_{n' \neq n, m} \left( v_{\alpha}^{nn'} L_{\gamma}^{n'm} + L_{\gamma}^{nn'} v_{\alpha}^{n'm} \right) \bigg] \Bigg\}.
\end{aligned} \label{S-OHC_CS}
\end{align}
\end{subequations}
Therefore the two contributions to the OHC will be:
\begin{equation}
    \sigma_{\alpha\beta}^{\gamma; D/ND}= -e\int_{BZ}\frac{d^3k}{(2\pi)^3}\sum_{n}f_n(\textbf{k})\Omega_{n,\alpha \beta}^{\gamma; D/ND}(\textbf{k}),
\end{equation}
from which we can write the total intrinsic contribution as
\begin{equation}
    \sigma_{\alpha\beta}^{\gamma; \hat{L}}=  \sigma_{\alpha\beta}^{\gamma; D} +  \sigma_{\alpha\beta}^{\gamma; ND}.
\end{equation}
Our results coincide with the ones in \cite{Sahu2021}.

\clearpage

\subsection{Weyl toy model for the field-induced OAM underlying the ND orbital-current response}

We introduce a minimal two-band Weyl toy model to isolate how an external electric field induces interband mixing between the two Weyl bands, generating a transverse OAM texture with the momentum-space symmetry required to contribute to the ND orbital-current response.

\textbf{Minimal Weyl Hamiltonian.} \\
We consider a single isotropic Weyl node \cite{Shen2017TopologicalInsulators, Vanderbilt2018BerryPhases},
\begin{equation}
\hat{H_0}(\mathbf{k})=\hbar v\,\mathbf{k}\cdot\bm{\hat{\tau}}
=\hbar v\,(k_x\tau_x+k_y\tau_y+k_z\tau_z),
\end{equation}
with eigenvalues
\begin{equation}
E_{\pm}(\mathbf{k})=\pm \hbar v k,
\qquad
k=\sqrt{k_x^2+k_y^2+k_z^2}.
\end{equation}
Introducing spherical coordinates,
\begin{equation}
\hat{\mathbf{k}}=(\sin\theta\cos\phi,\sin\theta\sin\phi,\cos\theta),
\end{equation}
the normalized eigenstates of $\hat{\mathbf{k}}\cdot\bm{\hat{\sigma}}$ can be chosen as
\begin{equation}
\lvert u_+(\mathbf{k})\rangle=
\begin{pmatrix}
\cos\frac{\theta}{2}\\[4pt]
 e^{i\phi}\sin\frac{\theta}{2}
\end{pmatrix} :=  \lvert u_c(\mathbf{k})\rangle,
\qquad
\lvert u_-(\mathbf{k})\rangle=
\begin{pmatrix}
-\sin\frac{\theta}{2}\\[4pt]
 e^{i\phi}\cos\frac{\theta}{2}
\end{pmatrix}:=  \lvert u_v(\mathbf{k})\rangle.
\label{S-Weyl spinors}
\end{equation}
They satisfy
\begin{equation}
\langle u_{\pm} \rvert \bm{\hat\sigma} \lvert u_{\pm} \rangle=\pm \hat{\mathbf{k}},
\end{equation}
so that the upper and lower bands have opposite helicity.

\textbf{Field-induced hybridization.}  \\
In the presence of a weak, spatially uniform electric field $\mathbf{\mathcal{E}}$, the adiabatically corrected instantaneous band state is obtained by treating the operator $\hat{H}(t)-i\hbar\partial_t$ to first order in the slow time dependence of $\mathbf{k}(t)$. Writing
\begin{equation}
\lvert \bar u_n \rangle=\lvert u_n \rangle+\lvert u_n^{(1)} \rangle,
\qquad
\bar E_n=E_n+E_n^{(1)},
\end{equation}
and inserting this expansion into
\begin{equation}
\left(\hat{H}-i\hbar\partial_t\right)\lvert \bar u_n \rangle=\bar E_n\lvert \bar u_n \rangle,
\end{equation}
one obtains, to first order,
\begin{equation}
(\hat{H}-E_n)\lvert u_n^{(1)} \rangle=
\left(i\hbar\partial_t+E_n^{(1)}\right)\lvert u_n \rangle.
\end{equation}
To determine the amplitude of the field-induced hybridization with other bands, we project this equation onto a state $\langle u_m|$ (where $m \neq n$). Because $\langle u_m | u_n \rangle = 0$, the first-order energy correction $E_{n}^{(1)}$ drops out. Operating $\hat{H}$ to the left yields its eigenvalue $E_m$, leading to:
\begin{equation}
    \langle u_{m}|u_{n}^{(1)}\rangle = \frac{i\hbar\langle u_{m}|\partial_{t}u_{n}\rangle}{E_{m}-E_{n}}.
\end{equation}
Using the chain rule $\partial_{t} = \dot{k}_{\alpha}\partial_{k_{\alpha}}$ -- the time dependence of the states enters solely through the evolution of the crystal momentum -- and substituting the semiclassical equation of motion $\hbar\dot{k}_{\alpha} = -e \mathcal{E}_{\alpha}$, the time derivative becomes $\partial_{t} = -\frac{e \mathcal{E}_{\alpha}}{\hbar}\partial_{k_{\alpha}}$. 

Then, summing the contributions from all other bands $m \neq n$, we arrive at the full first-order perturbed state:
\begin{equation}
\lvert \bar u_n \rangle=\lvert u_n \rangle+ie \mathcal{E}_{\alpha}\sum_{m\neq n}
\lvert u_m \rangle\frac{\langle u_m \rvert \partial_{k_{\alpha}} u_n \rangle}{E_n-E_m},
\label{S-eq:adiabatic_dressed_state_appendix}
\end{equation}
with repeated Cartesian indices understood. For the two-band Weyl cone one simply has
\begin{equation} 
\lvert \bar u_c \rangle=\lvert u_c\rangle+ie \mathcal{E}_{\alpha}
\lvert u_v \rangle\frac{\langle u_v \rvert \partial_{k_{\alpha}} u_c \rangle}{2\hbar vk},
\qquad
\lvert \bar u_v \rangle=\lvert u_v \rangle-ie \mathcal{E}_{\alpha}
\lvert u_c \rangle\frac{\langle u_c \rvert \partial_{k_{\alpha}} u_v \rangle}{2\hbar vk}.
\label{S-Weyl dressed states}
\end{equation}
Thus the applied field admixes the upper and lower cones with an amplitude controlled by the Berry connection matrix $\langle u_m \rvert \partial_{k_{\alpha}} u_n \rangle$ and inversely proportional to the band splitting $2\hbar vk$. Notice that Eqs. (\ref{S-Weyl dressed states}) are strictly valid in a close neighborhood of the node, but not exactly at the node.

\textbf{Generic first-order correction to an observable.}  \\
For any OAM component $\hat{L}_{\gamma}$, the first-order correction follows by evaluating the expectation value in the dressed state,
\begin{equation}
\delta\langle \hat{L}_{\gamma} \rangle_n
=\langle \bar u_n\rvert \hat{L}_{\gamma}\lvert \bar u_n\rangle-
\langle u_n\rvert \hat{L}_{\gamma}\lvert u_n\rangle.
\label{S-S24}
\end{equation}
By inserting the expansion $|\overline{u}_n\rangle = |u_n\rangle + |u_n^{(1)}\rangle$ into Eq. (\ref{S-S24}) and keeping only terms linear in the applied field, the zeroth-order term $\langle u_n|\hat{L}_{\gamma}|u_n\rangle$ cancels out. Neglecting the second-order contribution $\langle u_n^{(1)}|\hat{L}_{\gamma}|u_n^{(1)}\rangle$, the first-order correction simplifies to the sum of the cross terms:
\begin{equation}
    \delta\langle\hat{L}_{\gamma}\rangle_{n} \approx \langle u_{n}|\hat{L}_{\gamma}|u_n^{(1)}\rangle + \langle u_n^{(1)}|\hat{L}_{\gamma}|u_{n}\rangle.
\end{equation}

Exploiting the hermiticity of the orbital angular momentum operator ($\hat{L}_{\gamma}^\dagger = \hat{L}_{\gamma}$), the second term is simply the complex conjugate of the first, yielding:
\begin{equation}
    \delta\langle\hat{L}_{\gamma}\rangle_{n} = 2 \text{Re} \left[ \langle u_{n}|\hat{L}_{\gamma}|u_n^{(1)}\rangle \right].
\end{equation}

Substituting the explicit form of the first-order wave function correction $|u_n^{(1)}\rangle$ from Eq. (\ref{S-eq:adiabatic_dressed_state_appendix}), one directly obtains
\begin{equation}
\delta\langle \hat{L}_{\gamma} \rangle_n
=2\,\mathrm{Re}\sum_{m\neq n}
\left[
ie \mathcal{E}_{\alpha}
\frac{\langle u_n\rvert \hat{L}_{\gamma} \lvert u_m\rangle\langle u_m\rvert \partial_{k_{\alpha}} u_n\rangle}
{E_n-E_m}
\right].
\end{equation}
Using $2\,\mathrm{Re}(iz)=-2\,\mathrm{Im}(z)$, or equivalently reordering the matrix elements inside the numerator, this can be written as
\begin{equation}
\delta\langle \hat{L}_{\gamma} \rangle_n
=2e \mathcal{E}_{\alpha}\,\mathrm{Im}\!\left[
\sum_{m\neq n}
\frac{L_{\gamma}^{nm}\,\langle u_m \rvert \partial_{k_{\alpha}} u_n \rangle}
{E_n-E_m}
\right],
\label{S-eq:generic_deltaL_appendix}
\end{equation}
where 
\begin{equation}
L_{\gamma}^{nm}=\langle u_n \rvert \hat{L}_{\gamma} \lvert u_m \rangle.
\end{equation}
Eq. ~\eqref{S-eq:generic_deltaL_appendix} is a true interference formula, where the factor $\langle u_m \rvert \partial_{k_{\alpha}} u_n \rangle$ is geometric and the other,  $L_{\gamma}^{nm}$, is orbital. The induced OAM appears only when band geometry and orbital hybridization talk to each other. 
This happens when the \textit{non-diagonal} matrix elements of the OAM operator are non-zero:
\begin{equation}
L_{\gamma}^{nm}\neq 0.
\label{S-OHE_condition}
\end{equation}
This condition is naturally satisfied whenever the operator $\hat{L}_{\gamma}$ is not diagonal in the instantaneous band basis, which is generically expected in a multiorbital and strongly hybridized system.
For the two-band Weyl problem only the opposite band contributes, so for $n=v, c$ one gets
\begin{equation}
\delta\langle \hat{L}_{\gamma} \rangle_n
=2e \mathcal{E}_{\alpha}\,\mathrm{Im}\!\left[
\frac{L_{\gamma}^{n\bar n}\,\langle u_{\bar n} \rvert \partial_{k_{\alpha}} u_n \rangle}
{E_n-E_{\bar n}}
\right],
\label{S-eq:Weyl_deltaL_appendix}
\end{equation}
where $\bar n$ denotes the opposite band.

\textbf{Hall-like induced $\hat{L}_z$ for $\mathbf{\mathcal{E}}\parallel \hat{x}$.} \\

To verify that the induced OAM can contribute to a transverse orbital-current response, we examine its momentum-space parity for a Weyl cone.
To make the transverse structure explicit, consider a field along $x$, $\mathcal{E}=\mathcal{E}_x\hat{\mathbf{x}}$, and an OAM operator, in the Weyl bands subspace, of the form
\begin{equation}
\hat{L_z}=
\begin{pmatrix}
0 & \lambda_z\\
\lambda_z^* & 0
\end{pmatrix},
\qquad
\lambda_z=\lambda_R+i\lambda_I.
\label{S-S30}
\end{equation}
To evaluate the orbital polarization for both the conduction ($+$) and valence ($-$) bands, we introduce a band index $n = \pm 1$. The opposite band is denoted by $\bar{n} = -n$. The energy difference between the "active" and opposite band is $E_{n} - E_{\bar{n}} = 2n\hbar vk$. Using the Weyl spinors from Eq. (\ref{S-Weyl spinors}), the Berry connection matrix elements are:
\begin{equation}
    \langle u_{\bar{n}} | \partial_{k_{x}} u_{n} \rangle = n \frac{k_{x}k_{z}}{2k^{2}k_{\perp}} - i\frac{k_{y}}{2kk_{\perp}}.
\end{equation}
We parameterize the matrix elements of the hermitian OAM operator in the band basis in Eq. (\ref{S-S30}) by $L_{z}^{n\bar{n}} = \lambda_{R} + in\lambda_{I}$.
Substituting these generalized expressions into Eq. (\ref{S-eq:Weyl_deltaL_appendix}), the field-induced orbital angular momentum becomes:
\begin{equation}
    \delta\langle\hat{L}_{z}\rangle_{n} = 2e \mathcal{E}_{x} \text{Im} \left[ \frac{(\lambda_{R} + in\lambda_{I}) \left( n \frac{k_{x}k_{z}}{2k^{2}k_{\perp}} - i\frac{k_{y}}{2kk_{\perp}} \right)}{2n\hbar vk} \right].
\end{equation}
Multiplying the complex terms in the numerator and extracting the imaginary part, we find that the explicit dependence on $n$ within the imaginary part entirely factors out:
\begin{equation}
    \text{Im} \left[ (\lambda_{R} + in\lambda_{I}) \left( n \frac{k_{x}k_{z}}{2k^{2}k_{\perp}} - i\frac{k_{y}}{2kk_{\perp}} \right) \right] = \lambda_{I}\frac{k_{x}k_{z}}{2k^{2}k_{\perp}} - \lambda_{R}\frac{k_{y}}{2kk_{\perp}}.
\end{equation}
Pulling the remaining factor of $1/n = n$ to the front yields 
\begin{equation}
    \langle\hat{L}_{z}\rangle_{\pm} = \pm \frac{e\mathcal{E}_{x}}{2\hbar v} \left( -\lambda_{R}\frac{k_{y}}{k^{2}k_{\perp}} + \lambda_{I}\frac{k_{x}k_{z}}{k^{3}k_{\perp}} \right).
\end{equation}
The first term, coupled with $\lambda_{R}$, is odd under $k_{y}\rightarrow-k_{y}$, thus states with opposite transverse momentum $k_{y}$ carry opposite induced $\hat{L}_{z}$.
It is important to note that the physically relevant ingredient allowing for this transverse Hall-like response is not merely the real part of $\lambda_z$ , but the gauge-invariant product
\begin{equation}
\mathrm{Im}\!\left[\lambda_z(\mathbf{k})\,\langle u_v(\mathbf{k})\rvert \partial_{k_x}u_c(\mathbf{k})\rangle\right].
\label{S-eq:gauge_invariant_ingredient_appendix}
\end{equation}
Indeed, under a $U(1)$ rephasings $\lvert u_{\pm}\rangle\to e^{i\chi_{\pm}(\mathbf{k})}\lvert u_{\pm}\rangle$, one has
\begin{equation}
\lambda_z(\mathbf{k})\to e^{i[\chi_-(\mathbf{k})-\chi_+(\mathbf{k})]}\lambda_z(\mathbf{k}),
\qquad
\langle u_-\rvert \partial_{k_x}u_+\rangle\to e^{i[\chi_+(\mathbf{k})-\chi_-(\mathbf{k})]}\langle u_-\rvert \partial_{k_x}u_+\rangle,
\label{S-eq:gauge invariance}
\end{equation}
and therefore Eq.~\eqref{S-eq:gauge_invariant_ingredient_appendix} is unchanged. In a convenient gauge in which $\lambda_z$ is chosen real, the odd Hall-like part is isolated in the simplest possible form,
\begin{equation}
\langle \hat{L}_z \rangle_{\pm}^{\mathrm{odd}}
=\mp \frac{e\mathcal{E}_x\lambda_R}{2\hbar v}\frac{k_y}{k^2k_{\perp}}. \label{S-parity}
\end{equation}
This is the minimal toy-model realization of a momentum-resolved transverse orbital polarization generated by electric-field-induced interband hybridization in a Weyl semimetal.\\

\textbf{Condition for a non-zero ACA $L_{\gamma}^{nm}$ in a $d$-basis 2-bands system}\\
We now want to understand what are the physical conditions for Eq. (\ref{S-OHE_condition}) to be valid, i.e. for OHE to be driven by interband hybridization.
We assume $|u_{v}(\mathbf{k})\rangle$ and $|u_{c}(\mathbf{k})\rangle$ to be linear combinations of the $d$-basis functions, such that, for a given band index $n \in \{v,c\}$:
\begin{equation}
\lvert u_n(\mathbf{k}) \rangle= \alpha_n(\mathbf{k})\lvert d_{xy} \rangle + \beta_n(\mathbf{k})\lvert d_{yz} \rangle + \gamma_n(\mathbf{k})\lvert d_{z^2} \rangle + \delta_n(\mathbf{k})\lvert d_{xz} \rangle + \epsilon_n(\mathbf{k})\lvert d_{x^2-y^2} \rangle \\
:=\begin{pmatrix}
\alpha_n(\mathbf{k}) \\
\beta_n(\mathbf{k}) \\
\gamma_n(\mathbf{k}) \\
\delta_n(\mathbf{k}) \\
\epsilon_n(\mathbf{k})
\end{pmatrix}.
\label{S-eq:d-multiorbital}
\end{equation}
From now on, we fix $\mathbf{k}$ and omit the dependence on it. In this $d$-basis, the $\hat{L}_{z}$ operator in the ACA (Atomic center approximation) takes the form
\begin{equation}
    \hat{L}_{z} = \begin{pmatrix}
        0 & 0 & 0 & 0 & 2i \\
        0 & 0 & 0 & i & 0 \\
        0 & 0 & 0 & 0 & 0 \\
        0 & -i & 0 & 0 & 0 \\
        -2i & 0 & 0 & 0 & 0
    \end{pmatrix},
    \label{S-ACA-d-OAM}
\end{equation}
so that its diagonal elements in the $\{|u_{v}\rangle, |u_{c}\rangle\}$ subspace vanish. Its non-diagonal elements, on the other hand, which drive the field-induced OAM, are 
\begin{equation}
    \langle u_{n}|\hat{L}_{z}|u_{\bar{n}}\rangle = 2i(\alpha_{n}^{*}\epsilon_{\bar{n}} - \epsilon_{n}^{*}\alpha_{\bar{n}}) + i(\beta_{n}^{*}\delta_{\bar{n}} - \delta_{n}^{*}\beta_{\bar{n}}).
    \label{S-eq:ACA_multiorbital_codnition}
\end{equation}
Equation~(\ref{S-eq:ACA_multiorbital_codnition}) shows that a finite off-diagonal OAM matrix element arises when, at each $\mathbf{k}$-point, the two bands contain different complex orbital components in channels coupled by \(\hat L_z\), such as \(d_{xy}\leftrightarrow d_{x^2-y^2}\) or \(d_{yz}\leftrightarrow d_{xz}\). This provides the orbital ingredient required for the field-induced OAM, which can then feed the ND orbital Hall response. 

\end{document}